\documentclass[11pt,a4paper]{article}
\pdfoutput=1 
\usepackage[utf8]{inputenc}
\usepackage{tabularx}
\usepackage{lmodern}
\usepackage{mathtools}
\usepackage{longtable}
\usepackage{xspace}
\usepackage{cancel}
\usepackage{caption}
\usepackage{subcaption}
\usepackage[outercaption]{sidecap}
\usepackage{array} 
\usepackage{siunitx}
\usepackage{lscape}
\usepackage{microtype}
\usepackage{alpha}
\usepackage{macros}
\newcolumntype{C}{>{$}c<{$}}
\newcolumntype{R}{>{$}r<{$}}
\newcolumntype{L}{>{$}l<{$}}

\hypersetup{%
   pdftitle    = {The renormalised O(a) improved vector current
   	in three-flavour lattice QCD with Wilson quarks},
   pdfauthor   = {J. Heitger, F. Joswig},
   pdfkeywords = {Lattice QCD, Non-perturbative Effects,
                  Non-perturbative Renormalisation, Symanzik Improvement}
}%

\begin{document}

\preprintno{%
MS-TP-20-37\\
\vfill
}

\title{%
The renormalised $\Or(a)$ improved vector current\\
in three-flavour lattice QCD with Wilson quarks
}

\collaboration{\includegraphics[width=2.8cm]{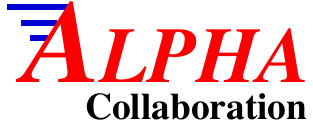}}

\author{Jochen Heitger and Fabian Joswig\,}

\address{Westf\"alische Wilhelms-Universit\"at M\"unster,
             Institut f\"ur Theoretische Physik,\\
             Wilhelm-Klemm-Stra{\ss}e 9, 48149 M\"unster, Germany}

\begin{abstract}
We present the results of a non-perturbative determination of the
improvement coefficient $\cv$ and the renormalisation factor $\zv$,
which define the renormalised vector current in three-flavour $\rmO(a)$
improved lattice QCD with Wilson quarks and tree-level Symanzik-improved
gauge action.
In case of the improvement coefficient, we consider both lattice
descriptions of the vector current, the local as well as the conserved
(i.e., point-split) one.
Our improvement and normalisation conditions are based on massive chiral
Ward identities and numerically evaluated in the Schr\"odinger functional
setup, which allows to eliminate finite quark mass effects in a controlled
way.
In order to ensure a smooth dependence of the renormalisation constant
and improvement coefficients on the bare gauge coupling, our computation
proceeds along a line of constant physics, covering the typical range of
lattice spacings $0.04\,\Fm\lesssim a\lesssim0.1\,\Fm$ that is useful for
phenomenological applications.
Especially for the improvement coefficient of the local vector current,
we report significant differences between the one-loop perturbative
estimates and our non-perturbative results.
\\
\end{abstract}

\begin{keyword}
Lattice QCD \sep Symanzik Improvement %
\sep Non-perturbative Effects \sep%
Non-perturbative Renormalisation 

\PACS{%
11.15.Ha\sep 
11.30.Rd\sep 
12.38.Gc\sep 
12.38.Aw     
}
\end{keyword}

\maketitle

\tableofcontents
\clearpage

\makeatletter
\g@addto@macro\bfseries{\boldmath}
\makeatother

\section{Introduction}
\label{sec:intro}
When formulating lattice QCD with Wilson fermions, its violation of
chiral symmetry is accompanied by the well-known characteristic that
the discretisation errors of the theory receive contributions linear
in the lattice spacing, $a$.
Despite the conceptual advantages of Wilson fermions\footnote{%
These include beneficial features such as strict locality, absence of
fermion doublers, and preservation of flavour symmetry as well as most
discrete symmetries.
}$\!\!$, as a practical shortcoming this implies a typically slow convergence
of physical quantities to the continuum limit. 
The obvious way to tame these linear discretisation errors
---~i.e., to approach the continuum limit by performing numerical
simulations at a series of decreasing, very small lattice spacings~---
generally still represents a computationally hard task, particularly in the
by now standard situation with three or four dynamical quark flavours.

According to the long-established recipe pioneered through Symanzik’s
continuum effective theory~\cite{Symanzik:1983dc,Symanzik:1983gh},
lattice artifacts can be systematically removed order by order in $a$ via
adding appropriate higher-dimensional operators to the action and to the
composite fields whose correlation functions are desired to extract
observables of interest in lattice QCD.
In fact, implementing this prescription by eliminating the leading,
$\rmO(a)$ cutoff effects has great computational impact, because it
accelerates the approach of physical quantities to the continuum in
numerical simulations with Wilson's lattice QCD.
If one restricts oneself to requiring improvement only for on-shell
quantities~\cite{Luscher:1984xn}, such as particle masses and matrix
elements between physical states, the structure of the improved action for
QCD and of the improved quark currents remains rather simple, see
Ref.~\cite{Luscher:1996sc} for a detailed early account on the general
theory.
More precisely, while for a removal of $\rmO(a)$ cutoff effects in
spectral quantities it suffices to improve the action by introducing the
dimension-five Sheikholeslami-Wohlert (aka, ``clover'')
term~\cite{Sheikholeslami:1985ij}, with a non-perturbatively fixed
coefficient $\csw$, the improvement of matrix elements of local operators
necessitates the addition of higher-dimensional counterterms to them,
together with the determination of associated improvement coefficients.
The present work employs the lattice discretisation of QCD with $\nf=3$
mass-degenerate flavours of Wilson quarks and the tree-level
Symanzik-improved gauge action~\cite{Luscher:1984xn};
the non-perturbatively tuned parameter $\csw$ to improve this fermion
action is available from Ref.~\cite{Bulava:2013cta}.

In tandem with the issue of cutoff effects and the rate of convergence of
the theory to the continuum limit, an unpleasant consequence of the
explicit breaking of all chiral symmetries by the Wilson term in the action
reflects in the Noether currents of chiral symmetry to be no longer
protected against renormalisation.
Therefore, the matrix elements of the axial and vector Noether currents
between hadron states and the vacuum, which are parameterised by
corresponding decay constants and constitute prime examples of physical
quantities that can be accurately measured in Monte Carlo simulations, need
to be correctly normalised.
The main approach to fix the current normalisation constants consists in
demanding continuum chiral symmetry relations to be fulfilled as
normalisation conditions at finite lattice
spacing~\cite{Bochicchio:1985xa,Luscher:1996jn}.
Since then, it has proven itself in numerous settings to implement this
strategy by exploiting chiral Ward identities, which follow from an
infinitesimal chiral change of variables in the QCD path integral and,
once evaluated numerically, are able to lead to a target precision that
does not dominate the uncertainty of the (renormalised) currents' matrix
elements.

In the following we focus on the non-perturbative improvement and
renormalisation of the (iso-vector flavour non-singlet) vector current,
since many interesting observables in lattice QCD are extracted from
correlation functions involving this current in one of its most widely used
discretisations, i.e., either the local vector current which has support
on only one lattice site, or the conserved point-split one which extends
over two neighbouring lattice points connected by a gauge link.
Analogously to the renormalised $\rmO(a)$ improved axial vector current,
which in the chiral limit of massless quarks is parameterised by an
improvement coefficient $\ca$ and a normalisation factor $\za$,
the improvement and renormalisation pattern of these two lattice vector
currents requires a single additive improvement term in each case,
whose respective coefficient $\cv$ gives the size of its $\rmO(a)$ mixing with the
(derivative of the) tensor current, multiplied by a finite
(i.e., renormalisation scale independent) normalisation factor $\zv$.
In the corresponding expressions, which for the vector channel are
explicitly given in the next section, improvement coefficients and
normalisations are functions of the bare coupling; $\ca$ and $\za$, are non-perturbatively known for our action
and phenomenologically relevant lattice spacings of about $0.1\,\Fm$ and
below from Refs.~\cite{Bulava:2015bxa} and
\cite{Bulava:2016ktf,DallaBrida:2018tpn}, respectively.

Our strategy to determine $\cv$ and $\zv$ non-perturbatively orients on
earlier investigations in the quenched
approximation~\cite{Guagnelli:1997db,Luscher:1996jn} and in two-flavour
QCD~\cite{DellaMorte:2005xgj} ($\zv$ only).
It adopts the Schrödinger functional formalism in conjunction with
exploiting massive chiral Ward identities, which relate correlation
functions of axial vector and vector currents.
In the $\rmO(a)$ improved theory, and for vanishing quark mass, these
identities can be written in a form that is valid up to error terms
quadratic in $a$.
Hence, the Schrödinger functional setup is particularly suited to
numerically evaluate improvement and normalisation conditions inferred from
Ward identities, because it enables to impose them in simulations close
to or even in the chiral limit, in the very spirit of a mass independent
scheme as it is adopted here.\footnote{%
Improvement conditions based on chiral Ward identities used in the past
in pure gauge theory without Schrödinger functional boundary conditions
can be found in Ref.~\cite{Bhattacharya:1999uq};
see also~\cite{Bhattacharya:2005rb} for its generalisation to lattice QCD
with light dynamical quarks of non-degenerate masses.
}
Previous non-perturbative determinations of improvement coefficients and
renormalisation factors, relying on the same Schrödinger functional gauge
field configuration ensembles with three flavours of mass-degenerate sea
quarks, include the aforementioned $\ca$~\cite{Bulava:2015bxa} and
$\za$~\cite{Bulava:2016ktf}, $b$-coefficients of additive quark mass terms
as well as (ratios of) $Z$-factors for $\rmO(a)$ improved quark mass
renormalisation~\cite{deDivitiis:2019xla,Heitger:2020mkp},
and the $\rmO(a)$ improvement of the tensor current~\cite{Chimirri:2019xsv}.
With the present vector channel study to determine $\cv$ and $\zv$ we
complement these works.
In the case of $\cv$, for instance, our computation goes beyond the
quenched one~\cite{Guagnelli:1997db} also methodigically, as it more
systematically includes, amongst a few more refinements on the technical
level, the finite quark mass contributions to eventually
eliminate them in a controlled way by an only slight extrapolation
(or even interpolation) of the results to the zero mass
limit in the unitary setup of equal sea and valence quark masses.

Other Ward identity calculations of $\zv$ for almost the same coupling
range of the three-flavour lattice QCD action at hand have recently been
published in Refs.~\cite{DallaBrida:2018tpn} and \cite{Gerardin:2018kpy},
but employ different frameworks.
While the former makes use of the so-called `chirally rotated' version of
the Schrödinger functional~\cite{Sint:2010eh,Brida:2016rmy}, the latter
directly operates on a subset of the large-volume $\nf=2+1$ gauge field
configuration ensembles with open boundary conditions, generated within the
joined effort by the CLS (Coordinated Lattice Simulations) cooperation of
lattice QCD
teams~\cite{Bruno:2014jqa,Bruno:2016plf,Bali:2016umi,Mohler:2017wnb}.
Moreover, \cite{Gerardin:2018kpy} also addresses the $\rmO(a)$ improvement
coefficients for the local and conserved vector currents discussed here
(and defined below), as well as the related $b$-coefficients for the
flavour non-singlet part of the local current that arise with the mass
dependence of the renormalisation factor if $\rmO(a)$ discretisation errors
are to be removed at physical values of the light quark masses.

In particular for the lattice prescription of the local current,
estimates of the relative contribution of the improvement term evaluated
with the perturbative one-loop value for the improvement coefficient $\cv$
known from~\cite{Taniguchi:1998pf} may suggest that the effect of
improvement in correlation functions entering, e.g., decay constants would
be small, and also the non-perturbative results found in
Ref.~\cite{Gerardin:2018kpy} stay very close to this perturbative
prediction, even in the region of stronger couplings.
On the other hand, already the preliminary outcome of the early quenched
calculation~\cite{Guagnelli:1997db} in the Schrödinger functional setup
indicated significant deviations of the non-perturbative estimates from
perturbation theory towards the strong-coupling regime.
And in fact, also in the present three-flavour investigation,
we observe sizeable values for $\cv$, which in the region
$g_0^2\gtrsim 1.55$ are far from one-loop perturbation theory, whereas we recover the expected asymptotics in agreement with
the perturbative behaviour when $g_0^2\to 0$.

Another conceptual difference compared to
Refs.~\cite{Guagnelli:1997db,Gerardin:2018kpy}
is that our improvement and renormalisation conditions based on chiral
Ward identities are imposed along a line of constant physics (LCP).
For this purpose, the spatial extent of the physical volume is fixed to
$L \approx 1.2\,\fm$, while $T/L\approx 3/2$, such that all length scales
in correlation functions are kept constant in physical units and only the
lattice spacing $a$ changes when $g_0$ is varied.\footnote{%
For an earlier discussion and applications of this idea,
see~\cite{Fritzsch:2010aw} and references therein.
}
Since this is just the situation, to which Symanzik's local effective
theory of cutoff effects can be applied, the resulting estimates of $\cv$
and $\zv$ are expected to exhibit a smooth dependence on the bare coupling.
Then it is obvious that any remaining intrinsic ambiguities, which could
potentially emerge through another choice for the LCP or, equivalently,
from a differently defined improvement or renormalisation condition, will
thus asymptotically disappear towards the continuum limit at a rate of at
least $\propto a$ in case of $c$-coefficients and $\propto a^2$ in case of
$Z$-factors if $\rmO(a)$ improvement is fully operative.

Since our range of bare gauge couplings covers lattice spacings
$0.04\,\Fm\lesssim a\lesssim0.1\,\Fm$, which matches those of the
aforementioned large-volume $\nf=2+1$ flavour QCD ensembles by CLS with
the same lattice action, our results may be used for a variety of
phenomenologically relevant applications involving the vector current,
for which knowledge of $\cv$ and $\zv$ is required.
To name a few prominent examples, these not only comprise computations of
vector meson decay constants, semi-leptonic decay form factors and the
leading-order hadronic contributions to the muon's anomalous magnetic
moment (a major part of which is accounted for by the timelike pion form
factor)\footnote{%
In computing these hadronic contributions, the electromagnetic (quark)
current plays a key r\^{o}le, which in QCD with the lightest three quark
flavours is conveniently written as a linear combination of two neutral
combinations from the octet of vector currents,
$V_\mu^{\text{el.-mag.}}=V_\mu^3+V_\mu^8/\sqrt{3}.$
},
but also thermal correlators for the di-lepton production rate in
the quark-gluon plasma.
As another potential area of usage we mention the non-perturbative matching
of Heavy Quark Effective Theory (HQET) to QCD in finite volume along the
lines of Refs.~\cite{Heitger:2003nj,DellaMorte:2013ega}, where the
renormalised axial and vector currents enter on the QCD side of the
matching calculation in order to extract the HQET parameters required for
determining B-meson decay constants and semi-leptonic form factors.

The remainder of this paper is structured as follows.
After a recapitulation of the theoretical framework
in Section~\ref{sec:theory}, in Section~\ref{sec:ensembles} we introduce the $\nf=3$
gauge field configurations underlying this work.
Section~\ref{sec:renormalisation} summarises the basic chiral Ward identities
between quark currents, our analysis, and final results
on the normalisation constant $\zv$ of the vector current and compares them
with those of Refs.~\cite{DallaBrida:2018tpn,Gerardin:2018kpy}. 
As a byproduct, we also obtain the coefficient of the (valence) quark mass
dependent piece of the renormalisation factor away from the chiral limit,  
$b_\mathrm{V}$, for one value of the lattice spacing to demonstrate
consistency with the results of~\cite{Fritzsch:2018zym,Gerardin:2018kpy}.
The Ward identity determination of the improvement coefficients for both, the local as
well as the conserved (i.e., point-split) lattice discretisation of the
continuum vector current is discussed in Section~\ref{sec:improvement},
finally yielding a continuous parameterisation of our non-perturbatively
computed values as a function of the bare coupling.
Since in the strong-coupling domain and especially for $\cv$ we find,
by contrast to Ref.~\cite{Gerardin:2018kpy}, sizable non-perturbative
corrections to the one-loop formula, we present a few consistency checks
that demonstrate higher-order ambiguities among these different
determinations to vanish as $g_0^2\to 0$.
Section~\ref{sec:concl} contains our conclusions; further data tables and definitions of Schrödinger functional correlation functions are deferred to appendices.

A preliminary account on the present work was reported
in~\cite{Heitger:2017njs}.

\section{Definitions}
\label{sec:theory}
To prepare the ground for the non-perturbative improvement and
renormalisation of the vector current in Wilson's lattice QCD through
numerical simulations, this section summarises the theoretical framework and a few relevant definitions.
\subsection{Quark currents}
We define the flavour non-singlet fermion bilinears associated with the vector current, tensor current, axial vector current, and the pseudoscalar density as 
\begin{align}
V_\mu^a(x)&=\bar{\psi}(x)T^a\gamma_\mu\psi(x)\,, &&T_{\mu\nu}^a(x)=\mathrm{i}\bar{\psi}(x)T^a\sigma_{\mu\nu}\psi(x)\,,\\
A_\mu^a(x)&=\bar{\psi}(x)T^a\gamma_\mu\gamma_5\psi(x)\,, &&\,\,P^a(x)=\bar{\psi}(x)T^a\gamma_5\psi(x)\,,
\end{align}
where $T^a$ are the anti-hermitean generators of $\mathrm{SU}(N_\mathrm{f})$. In this work we consider three flavours of degenerate quarks with the bare subtracted quark mass
\begin{align}
m_{\mathrm{q}}=m_0-m_\mathrm{crit}\,.
\end{align}
The corresponding current quark mass can be obtained via the PCAC relation as
\begin{align}
m = \frac{\partial_0\langle A_0^a(x) \mathit{O}^a\rangle}{2\langle P^a(x) \mathit{O}^a\rangle}\,,
\end{align}
for any operator $\mathit{O}^a$.
In applications to hadronic physics one would also like to control the
$\rmO(a)$ effects that according to Symanzik's programme applied to
Wilson's lattice QCD are canceled by the counterterms to the currents.
For the axial and vector currents, and close to the chiral limit,
one essentially requires counterterms with coefficients $c_\mathrm{A}$ and $c_\mathrm{V}$, respectively.
More precisely, the improvement and renormalisation pattern of these
currents reads
\begin{align}
(V_\mathrm{R}^\mathrm{I})_\mu^a&=Z_\mathrm{V}\big(1+b_\mathrm{V}am_\mathrm{q}+\bar{b}_\mathrm{V}\mathrm{Tr}[aM_\mathrm{q}]\big)\big(V_\mu^a+ac_\mathrm{V}\tilde{\partial}_\nu T_{\mu\nu}^a\big)\,,\\
(A_\mathrm{R}^\mathrm{I})_\mu^a&=Z_\mathrm{A}\big(1+b_\mathrm{A}am_\mathrm{q}+\bar{b}_\mathrm{A}\mathrm{Tr}[aM_\mathrm{q}]\big)\big(A_\mu^a+ac_\mathrm{A}\tilde{\partial}_\mu P^a\big)\,,
\end{align}
where $M_\mathrm{q}=\operatorname{diag}(m_{\mathrm{q},1},m_{\mathrm{q},2},m_{\mathrm{q},3})=m_\mathrm{q}\mathbf{1}$.

Alternatively one can define a conserved vector current, which is not local but split among neighboring lattice sites:
\begin{align}
\hat{V}_\mu^{a}(x)=\frac{1}{2}\Big( \bar{\psi}(x+a\hat{\mu})T^a(1+\gamma_\mu)U_\mu^\dagger(x)\psi(x)-\bar{\psi}(x)T^a(1-\gamma_\mu)U_\mu(x)\psi(x+a\hat{\mu}) \Big)\,,
\end{align}
In practice we employ a symmetrised version of the conserved vector current \cite{Frezzotti:2001ea} defined as
\begin{align}
\tilde{V}_\mu^{a}(x)=&\frac{1}{2}\Big(\hat{V}_\mu^{a}(x)+\hat{V}_\mu^{a}(x-a\hat{\mu})  \Big)\,,
\end{align}
which under spacetime reflections transforms in the same way as the local vector current.\footnote{In all situations where we consider the conserved vector current, we explicitly sum over all spatial lattice points. Due to periodic boundary conditions and zero momentum, the spatial components of the two variants of the conserved current are equivalent when the sum is taken: $	\sum_{\mathbf{x}}\hat{V}_k^a(x_0,\mathbf{x})=\sum_{\mathbf{x}}\tilde{V}_k^a(x_0,\mathbf{x})$.}
As the name suggests, the conserved vector current does not require renormalisation but local improvement via a counterterm parametrised by the coefficient $c_{\tilde{\mathrm{V}}}$
\begin{align}
(\tilde{V}_\mathrm{R}^\mathrm{I})_\mu^a=\tilde{V}_\mu^a+ac_{\tilde{\mathrm{V}}}\tilde{\partial}_\nu T_{\mu\nu}^a\,.
\end{align}

For the renormalisation of the local vector current and the accompanying improvement coefficient there exist one-loop perturbative predictions for the tree-level Symanzik-improved
gauge action \cite{Aoki:1998ar, Taniguchi:1998pf}:
\begin{align}
Z_\mathrm{V}(g_0^2)&=1-0.075427C_\mathrm{F}g_0^2+\mathrm{O}(g_0^4)\,,\\
c_\mathrm{V}(g_0^2)&=-0.01030(4)C_\mathrm{F}g_0^2+\mathrm{O}(g_0^4)\,.
\end{align}
For the improvement coefficient of the conserved vector current, $c_{\tilde{\mathrm{V}}}$, only its tree-level value $\tfrac{1}{2}$ is known.
\subsection{Schrödinger functional}
For the numerical determination of the renormalisation and improvement factors of interest we evaluate the QCD path integral on a hypercubic space-time lattice with Schrödinger functional boundary conditions (i.e., periodic in space and Dirichlet in time). Since these introduce a gap into the spectrum of the Dirac operator, the use of the Schrödinger functional enables us to simulate with quark masses very close to the chiral point defined by $am=0$. Another advantage can be gained by exploiting the property that the boundary time slices are invariant under global gauge transformations. 
The parity constraints at the boundary only allow for fermion bilinears that are parity odd. This essentially leaves us with two choices for the boundary sources, one corresponding to a pseudoscalar, the other to a vector operator:
\begin{align}
\mathcal{O}^a=a^6\sum_{\mathbf{u},\mathbf{v}}\bar{\zeta}(\mathbf{u})T^a\gamma_5\zeta(\mathbf
v)\,,\quad \mathcal{O}^{\prime a}=a^6\sum_{\mathbf{u},\mathbf{v}}\bar{\zeta}^\prime(\mathbf{u})T^a\gamma_5\zeta^\prime(\mathbf
v)\,,\\
\mathcal{Q}^a_k=a^6\sum_{\mathbf{u},\mathbf{v}}\bar{\zeta}(\mathbf{u})T^a\gamma_k\zeta(\mathbf
v)\,,\quad \mathcal{Q}^{\prime a}_k=a^6\sum_{\mathbf{u},\mathbf{v}}\bar{\zeta}^\prime(\mathbf{u})T^a\gamma_k\zeta^\prime(\mathbf
v)\,.
\end{align}
The explicit forms of the Schrödinger functional boundary-to-bulk and boundary-to-boundary correlation functions relevant for constructing the estimators of improvement coefficients and renormalisation factors computed here are given in later sections and in Appendix~\ref{sec:sf_correlation_functions}, respectively.

\section{Gauge field ensembles and analysis aspects}
\label{sec:ensembles}
The parameters of the gauge field configuration ensembles used in this study are summarised in Table~\ref{tab:gauge_parameters}. The ensembles are designed to lie on a line of constant physics (LCP) defined by a physical extent $L\approx1.2\,\Fm$. As the original tuning was based on the perturbative (universal) two-loop beta-function as explained in \cite{Bulava:2015bxa}, $L$ is only constant up to $\mathrm{O}(g_0^2)$ effects. We explicitly confirmed by exploiting the results on the scale setting for our lattice action from \cite{Bruno:2016plf} that these cutoff effects are consistent with $\mathrm{O}(a)$ (cf.\ discussion in Subsection~\ref{subsec:lcp_violation}). We thus expect our final results to be only affected at $\mathrm{O}(a^2)$ for renormalisation constants and $\mathrm{O}(a)$ for improvement coefficients, which is beyond the order we are interested in and treated as an ambiguity that extrapolates to zero in the continuum limit and is not accounted for in the final error budget.
Further details on the numerical simulations to generate the gauge field ensembles in Table~\ref{tab:gauge_parameters} can be found in refs.~\cite{Bulava:2015bxa,Bulava:2016ktf,Heitger:2017njs,deDivitiis:2019xla,Chimirri:2019xsv,Heitger:2020mkp}, where they have already been the basis of previous work on non-perturbative improvement and renormalisation in three-flavour QCD by means of imposing Ward identities formulated in terms of Schr\"odinger functional correlation functions.
\begin{table}[t]
	\centering
	\renewcommand{\arraystretch}{1.25}
	\setlength{\tabcolsep}{3pt}
\begin{tabular}{llllrcl}
	\toprule
	ID       & $L^3\times T/a^4$   &   $\beta$ &   $\kappa$ &   MDU &   $P(Q=0)$ & $\tau_\mathrm{exp}$   \\
	\midrule
 A1k1     & $12^3\times 17$     &     3.3   &   0.13652  & 20480 &    0.365 & 1.041(58)             \\
A1k3     & $12^3\times 17$     &     3.3   &   0.13648  &  6876 &    0.357 & 2.08(12)              \\
A1k4     & $12^3\times 17$     &     3.3   &   0.1365   & 96640 &    0.366 & 1.041(58)             \\
\midrule
E1k1     & $14^3\times 21$     &     3.414 &   0.1369   & 38400 &    0.353 & 1.28(10)              \\
E1k2     & $14^3\times 21$     &     3.414 &   0.13695  & 57600 &    0.375 & 1.28(10)              \\
\midrule
B1k1     & $16^3\times 23$     &     3.512 &   0.137    & 20480 &    0.389 & 2.78(41)              \\
B1k2     & $16^3\times 23$     &     3.512 &   0.13703  &  8192 &    0.341 & 2.78(41)              \\
B1k3     & $16^3\times 23$     &     3.512 &   0.1371   & 16384 &    0.458 & 2.78(41)              \\
B1k4     & $16^3\times 23$     &     3.512 &   0.13714  & 27856 &    0.402 & 2.78(41)              \\
\midrule
C1k1     & $20^3\times 29$     &     3.676 &   0.1368   &  7848 &    0.334 & 7.9(2.1)              \\
C1k2     & $20^3\times 29$     &     3.676 &   0.137    & 15232 &    0.45  & 7.9(2.1)              \\
C1k3     & $20^3\times 29$     &     3.676 &   0.13719  & 15472 &    0.645 & 7.9(2.1)              \\
\midrule
D1k2     & $24^3\times 35$     &     3.81  &   0.13701  &  6424 &    0.457 & 19.2(3.9)             \\
D1k4     & $24^3\times 35$     &     3.81  &   0.137033 & 85008 &    0.696 & 9.6(1.9)              \\
	\bottomrule
\end{tabular}
	\caption{Summary of simulation parameters for the gauge configuration ensembles labeled by `ID'. MDU denotes the total number of molecular dynamics units. $P(Q_0)$ labels the percentage of configurations, for which the topological charge $Q$ vanishes. $\tau_\mathrm{exp}$ indicates the exponential autocorrelation time used for the tail within the data analysis along~\cite{Schaefer:2010hu}. It is estimated from the integrated autocorrelation time of the correlator $F_1$ on the longest Monte Carlo chain for each value of $\beta$. All configurations in a given ensembles are separated by $8$ molecular dynamics units, except for ensembles A1k3 (4) and D1k4 (16). This is why $\tau_\mathrm{exp}$ for these ensembles differs from the others at the same value of $\beta$.}
	\label{tab:gauge_parameters}
\end{table}

In numerical simulations by standard Monte Carlo algorithms (and if the physical volume is not so small that fluctuations of topology are a priori suppressed) it becomes increasingly difficult to sufficiently sample all topological sectors of QCD towards the continuum limit.
This problem, often referred to as ``topology freezing'', implies that for the finer lattice spacings our simulations and the results extracted from them may suffer from critical slowing down of $Q$.
To make sure that an incorrect sampling of the topological charge does not affect our data, we project expectation values to the topologically trivial sector by reweighting, as suggested in~\cite{Fritzsch:2013yxa,DallaBrida:2016kgh} and also applied in Refs.~\cite{Bulava:2015bxa,Bulava:2016ktf,DallaBrida:2018tpn,deDivitiis:2019xla,Heitger:2020mkp,Heitger:2017njs,Chimirri:2019xsv} in similar contexts before. Circumventing this algorithmic issue by the restriction to the trivial topological sector provides a viable and theoretically sound procedure, since it corresponds to imposing within the $Q=0$ sector improvement and renormalization conditions, which derive from chiral flavour symmetries holding separately in each topological charge sector. For the definition to evaluate the topological charge, we follow the prescription adopted in Refs.~\cite{Fritzsch:2013yxa,DallaBrida:2016kgh}. While this projection to $Q=0$ hence comes at the expense of typically somewhat larger statistical errors\footnote{For our ensembles, the percentage of gauge field configurations with $Q\neq 0$ is generally around $\approx 60\%$ and decreases to $\approx 30\%$ towards the finest lattice spacing, see Table~\ref{tab:gauge_parameters}. This reduction of statistics, however, is partly compensated by an associated decorrelation of the data.}$\!$ and a slightly modified cutoff dependence, it is expected (and also confirmed in our data) to not induce a noticeable difference in the final results and their $g_0^2$-dependence exhibiting a smooth approach to the continuum limit.

For the error analysis we use (a python implementation of) the $\Gamma$-method~\cite{Wolff:2003sm}, also accounting for effects of critical slowing down according to~\cite{Schaefer:2010hu}, incorporating the techniques of automatic differentiation as suggested in \cite{Ramos:2018vgu}. As estimate for the autocorrelation time $\tau_\mathrm{exp}$ of the slowest mode in a given Monte Carlo history (after $Q=0$ projection), we employ the integrated autocorrelation time of the boundary-to-boundary correlation function $F_1$ defined in eq.~(\ref{eq:definition_F1}), based on the assumption that, due to the parity invariance of the HMC algorithm, it is sufficient to look for the slowest mode in parity-even observables \cite{Schaefer:2010hu}. We determine $\tau_\mathrm{exp}$ for the longest Monte Carlo chain available for each value of $\beta$ and conservatively consider this result as also valid for the remaining ensembles, see Table~\ref{tab:gauge_parameters}. Only in a few cases, $\tau_\mathrm{exp}$ has to be adjusted by a factor stemming from different configuration spacings in molecular dynamics units (MDU). 

\section{Renormalisation of the vector current}
\label{sec:renormalisation}
We estimate the finite renormalisation of the vector current by demanding that the vector Ward identity in the $\Or(a)$ improved theory is fulfilled up to $\Or(a^2)$ corrections. In the present three-flavour case, this idea is numerically implemented within the finite-volume Schr\"odinger functional setup, similarly to previous calculations in the quenched approximation and in two-flavour QCD~\cite{Luscher:1996jn,DellaMorte:2005xgj}, while here in addition we also explore to fix the normalisation of the vector current via an alternative choice of boundary operators, yielding another non-perturbative estimate of $\zv$ with different  $\Or(a^2)$ corrections.
Other recently reported determinations of $\zv$ in the three-flavour theory for the same discretisation and almost the same $\beta$-range employ the chirally rotated Schr\"odinger functional~\cite{DallaBrida:2018tpn} and the large-volume gauge field ensembles with open boundary conditions generated by CLS~\cite{Gerardin:2018kpy}.
\subsection{Renormalisation condition and its evaluation}
We consider infinitesimal vector rotations of the quark fields defined as
\begin{align}
\delta_\mathrm{V}^a\bar\psi(x)\approx-\mathrm{i}\bar\psi(x)T^a\,,\quad 
\delta_\mathrm{V}^a\psi(x)\approx\mathrm{i}T^a\psi(x)\,.
\end{align}
For the present discussion we restrict ourselves to the $\SUtwo$ subgroup of $\SUthree$.
Based on the invariance of composite operators under these rotations,
\begin{align}
\big\langle \mathit{O}\,\delta_\mathrm{V}S\big\rangle = \big\langle \delta_\mathrm{V}\mathit{O}\big\rangle \,,
\end{align}
one can derive the vector Ward identity in an integrated form:
\begin{align}
\int \mathrm{d}^3\mathbf{x} \,\big\langle \mathit{O}^{\prime c}(z)\, V_0^a(x_0,\mathbf{x}) \mathit{O}^b(y)\big\rangle = -\,\big\langle \mathit{O}^{\prime c}(z) [\delta_\mathrm{V}^a\mathit{O}^{b}(y)] \big\rangle \,,
\end{align}
where we limit the vector variation to all time slices where $y_0<x_0$ and set $z_0>x_0$.
We identify the two operators $\mathit{O}$ and $\mathit{O}^\prime$ with Schrödinger functional boundary interpolators, which only change their flavour structure but preserve their Dirac structure under small vector rotations
\begin{align}
\delta_\mathrm{V}^a\mathcal{O}^b(x)=-\mathrm{i}\epsilon^{abc}\mathcal{O}^c(x)\,,\quad \delta_\mathrm{V}^a\mathcal{Q}_k^b(x)=-\mathrm{i}\epsilon^{abc}\mathcal{Q}_k^c(x)\,,
\end{align}
with $\epsilon^{abc}$ the totally anti-symmetric tensor of $\SUtwo$.
Utilizing the particular types of Schrödinger functional boundary-to-boundary correlation functions defined as
\begin{align}
F_\mathrm{V}(x_0)&=-\frac{\mathrm{i}a^3}{2L^6}\epsilon^{abc}\sum_{\mathbf{x}}\big\langle \mathcal{O}^{\prime a}V_0^b(x_0,\mathbf{x})\mathcal{O}^c\big\rangle\,,\quad \,F_1=-\frac{1}{2L^6}\big\langle \mathcal{O}^{\prime a}\mathcal{O}^a\big\rangle\,,\label{eq:definition_F1}\\
K_\mathrm{V}(x_0)&=-\frac{\mathrm{i}a^3}{6L^6}\epsilon^{abc}\sum_{\mathbf{x}}\big\langle \mathcal{Q}_k^{\prime a}V_0^b(x_0,\mathbf{x})\mathcal{Q}_k^c\big\rangle\,,\quad K_1=-\frac{1}{2L^6}\big\langle \mathcal{Q}_k^{\prime a}\mathcal{Q}_k^a\big\rangle\,,
\end{align}
we arrive at two variants of the vector Ward identity, viz.
\begin{align}
Z_\mathrm{V}^\mathrm{f}\big(1+b_\mathrm{V}^\mathrm{f}am_\mathrm{q}+\bar{b}_\mathrm{V}^\mathrm{f}\mathrm{Tr}[aM_\mathrm{q}]\big)&=\frac{F_1}{F_{\mathrm{V}}(x_0)}\,+\mathrm{O}(a^2)\,,
\label{eq:def_ZVf}\\
Z_\mathrm{V}^\mathrm{k}\big(1+b_\mathrm{V}^\mathrm{k}am_\mathrm{q}+\bar{b}_\mathrm{V}^\mathrm{k}\mathrm{Tr}[aM_\mathrm{q}]\big)&=\frac{K_1}{K_{\mathrm{V}}(x_0)}+\mathrm{O}(a^2)\,,
\label{eq:def_ZVk}
\end{align}
from which $Z_\mathrm{V}$ can readily be extracted.
Since in the spirit of adopting a quark mass independent renormalisation scheme we ultimately work in the (unitary) limit of vanishing degenerate sea and valence quark masses, the terms proportional to $am_\mathrm{q}$ and $\mathrm{Tr}[aM_\mathrm{q}]$ could have already been dropped at this stage.

\begin{figure}[t]
	\centering
	\begin{subfigure}{.5\textwidth}
		\centering
		\includegraphics[width=\linewidth]{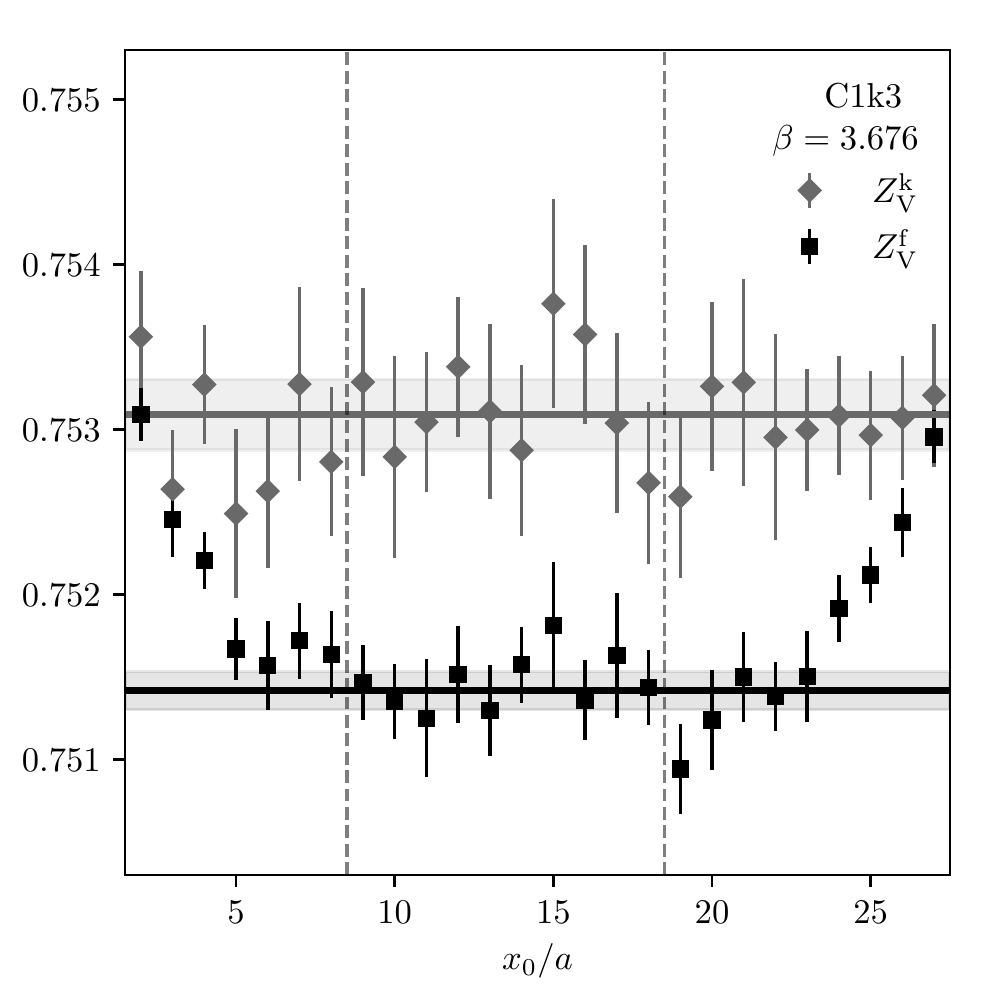}
	\end{subfigure}%
	\begin{subfigure}{.5\textwidth}
		\centering
		\includegraphics[width=\linewidth]{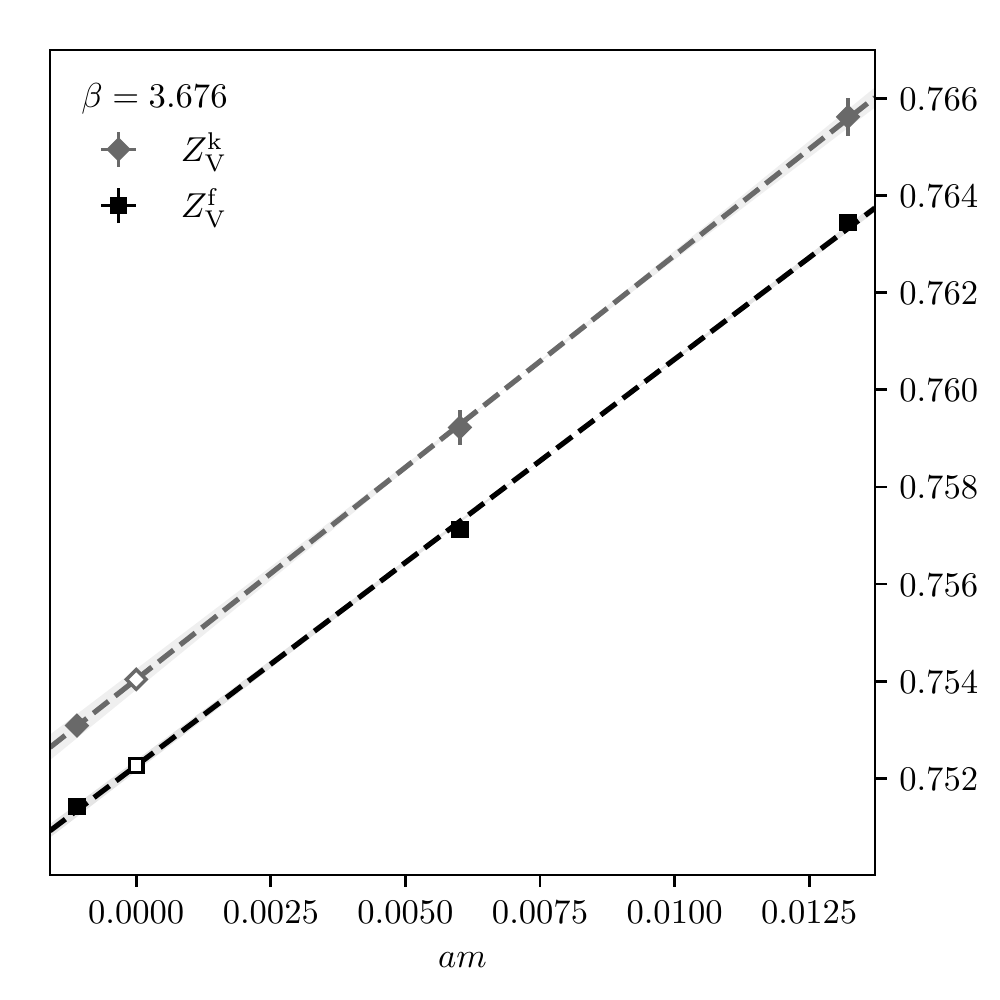}
	\end{subfigure}
	\caption{{\em Left:} Plateaux of the two determinations of $Z_\mathrm{V}$ for ensemble C1k3 ($\beta=3.676$). The dashed vertical lines enclose the plateau region, the horizontal lines with shaded regions correspond to the plateau values and their uncertainties. {\em Right:} Chiral extrapolation of the two variants of definitions of $Z_\mathrm{V}$ for $\beta=3.676$. The dashed lines correspond to linear fits to the data, while the shaded regions describe the uncertainties of the fits. Open symbols mark the respective value in the chiral limit. Note that, thanks to the Schr\"odinger functional boundary conditions, even (slightly) negative PCAC quark masses are possible such that in those cases (as illustrated here) extrapolations actually become interpolations.}
	\label{fig:ZV_extraction}
\end{figure}

We evaluate these two different variants of the Ward identity for every time slice and identify a plateau as depicted in the left part of Figure~\ref{fig:ZV_extraction}. As for the plateau average, we decide to use the central third of the temporal extent of the lattice such that the plateau lengths are ensured to be kept approximately constant in physical units. Generally, the variant $Z_\mathrm{V}^\mathrm{k}$ seems to be less sensitive to boundary affect and settles to a plateau at smaller distances from the temporal boundary. However, the statistical fluctuations are smaller by a factor $\approx 2$ in the case of $Z_\mathrm{V}^\mathrm{f}$, which is why we prefer this definition for our final results. We explictly confirm below that settling on one of these two definitions of the renormalisation constant only amounts to ambiguities which vanish with a rate $\propto a^2$ as expected in the $\Or(a)$ improved theory.

As illustrated for representative examples in the right part of Figure~\ref{fig:ZV_extraction}, a linear function is perfectly justified to extract the renormalisation constants defined at zero quark mass\footnote{Here and in all further chiral extrapolations we employ the PCAC definition of the quark mass.}$\!\!$ from the data at given bare coupling under investigation.
Here, the use of Schrödinger functional boundary condition facilitates very small positive and even negative quark masses, which then sometimes promotes the chiral extrapolations to interpolations and thus renders the fits very reliable. For the fitting procedure we rely on the orthogonal distance regression algorithm~\cite{Boggs1989}, which takes into account errors in both the dependent and independent variables. This increases the statistical error of the value at the chiral point by up to $20$\% in comparison to the standard least-squares procedure. As argued above, we do not use any $b$-coefficients in the chiral extrapolation, because our simulations are sufficiently close to the chiral limit. The results for the individual ensembles and its chiral limits are summarised in Table~\ref{tab:ZV_results}.

\begin{figure}[th!]
	\centering
	\begin{subfigure}{.5\textwidth}
		\centering
		\includegraphics[width=\linewidth]{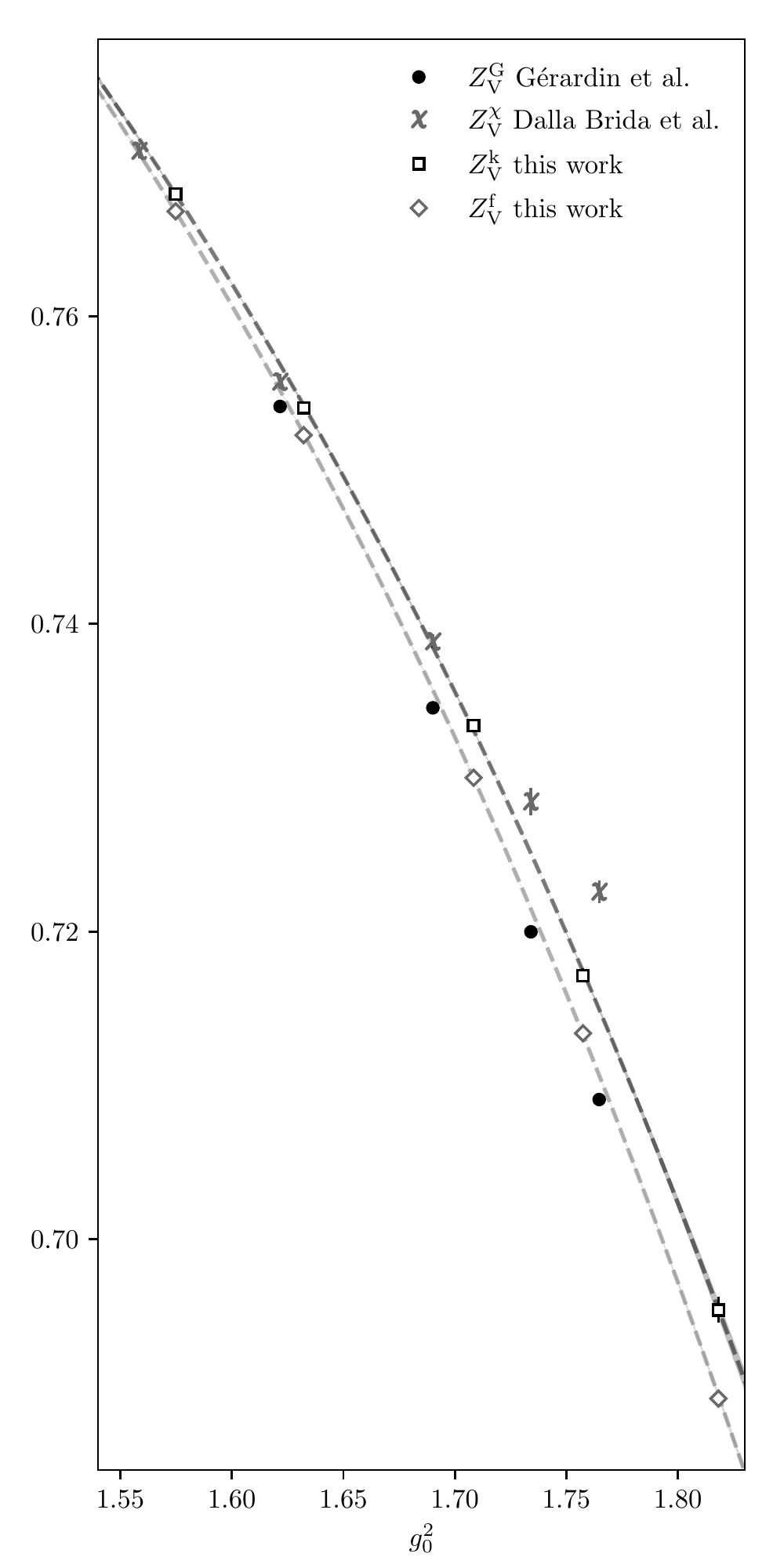}
	\end{subfigure}%
	\begin{subfigure}{.5\textwidth}
		\centering
		\includegraphics[width=0.997\linewidth]{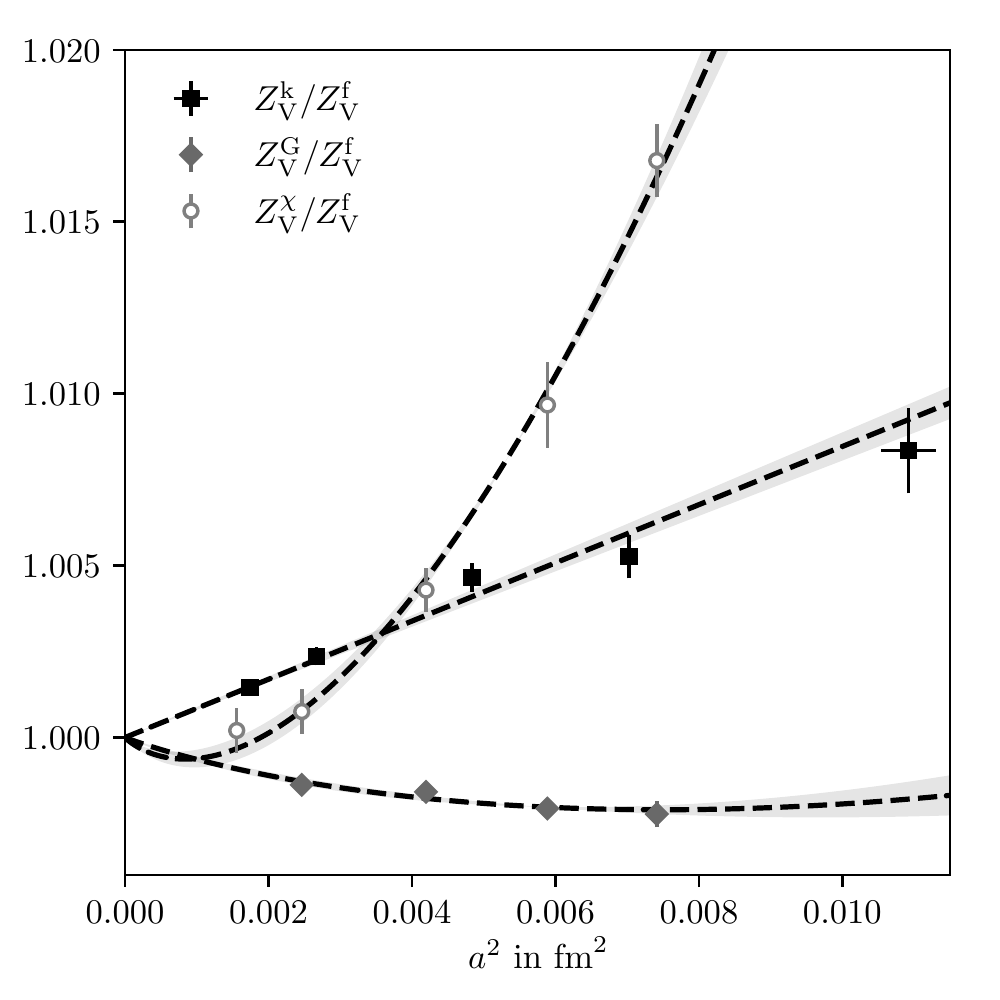}
		\includegraphics[width=0.997\linewidth]{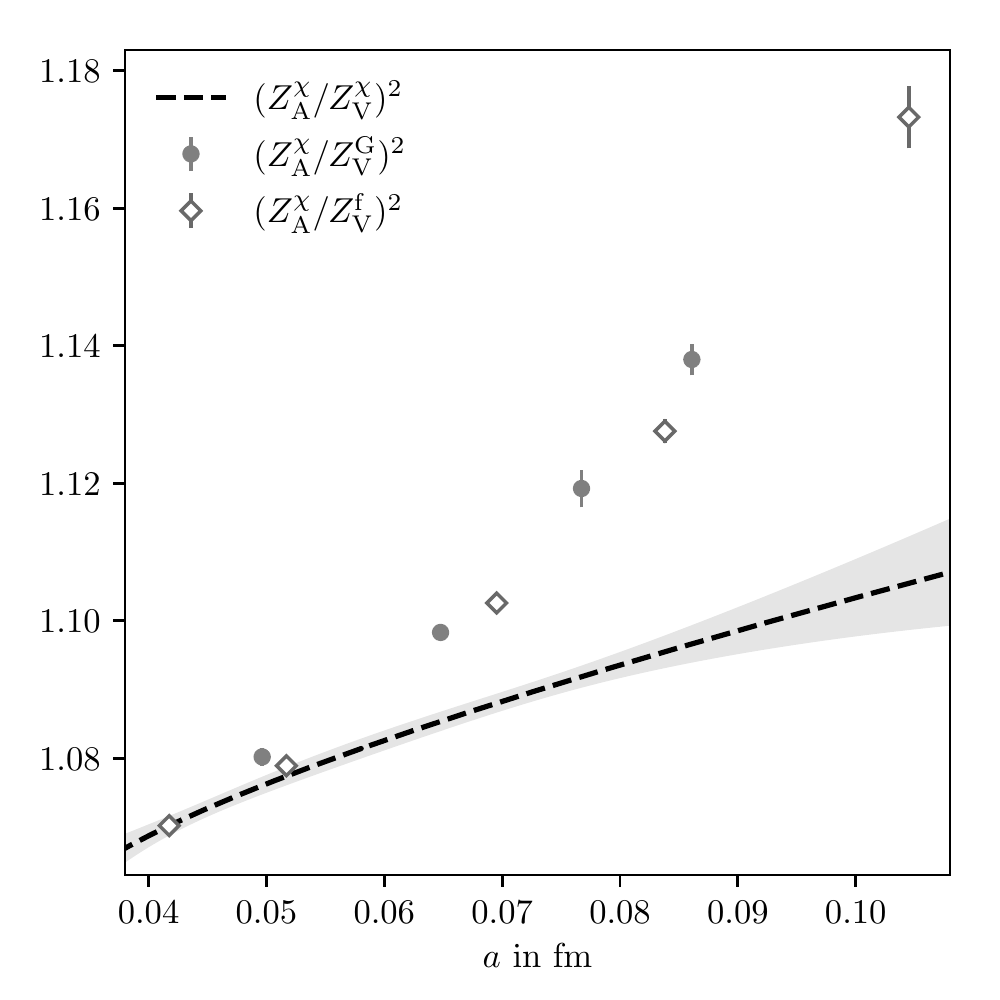}
	\end{subfigure}
	\caption{{\em Left:} Bare coupling dependence of the two variants of $Z_\mathrm{V}$ in comparison to the results of \cite{DallaBrida:2018tpn,Gerardin:2018kpy}. The dashed lines and the surrounding shaded areas correspond to the interpolation formulae and the respective uncertainties as described in the text. For future applications we advocate the use of $Z_\mathrm{V}^\mathrm{f}$. {\em Top right:} Continuum limit of the ratio of various determinations of $Z_\mathrm{V}$ (from this work and the literature) with our preferred definition $Z_\mathrm{V}^\mathrm{f}$. The dashed lines describe the functional forms which best describe the lattice spacing dependence as detailed in the text. {\em Bottom right:} Continuum limit of the squared ratio of $Z_\mathrm{A}^\chi$ and various determinations of $Z_\mathrm{V}$.}
	\label{fig:ZV_overview}
\end{figure}

In the left part of Figure~\ref{fig:ZV_overview}, the $g_0^2$-dependences of our two determinations are presented in comparison to the results of Refs.~\cite{DallaBrida:2018tpn,Gerardin:2018kpy}.
Our non-perturbative results can very well be parametrised by smooth interpolation formulae of a form asymptotically approaching the one-loop perturbative prediction~\cite{Aoki:1998ar} in the $g_0^2\to 0$ limit,
\begin{subequations}
\label{eq:interpolation_ZV}
\begin{align}
Z_\mathrm{V}^\mathrm{x}(g_0^2)=1-0.075427C_\mathrm{F}g_0^2+Z_\mathrm{V}^{\mathrm{x},(0)}g_0^4+Z_\mathrm{V}^{\mathrm{x},(1)}g_0^6+Z_\mathrm{V}^{\mathrm{x},(2)}g_0^8 \,,
\end{align}
where, as discussed before, we advocate in future applications to employ the estimator corresponding to $\mathrm{x}=\mathrm{f}$, with fitted coefficients and covariance matrix 
\begin{align}
	Z_\mathrm{V}^{\mathrm{f},(i)} &=	
	\begin{pmatrix}
	\begin{tabular}{@{}*{1}{S[table-format = +3.3e+1]}}
	-1.907e-01 \\
	+2.199e-01 \\
	-7.492e-02
	\end{tabular}
	\end{pmatrix} \,, \\
	\operatorname{cov}(Z_\mathrm{V}^{\mathrm{f},(i)})&=
	\begin{pmatrix}
	\begin{tabular}{@{}*{3}{S[table-format = +3.5e+1]}}
	+1.11887e-04 & -1.34335e-04 & +4.02275e-05 \\
	-1.34335e-04 & +1.61288e-04 & -4.82986e-05 \\
	+4.02275e-05 & -4.82986e-05 & +1.44633e-05 
	\end{tabular}
	\end{pmatrix} \,.
\end{align}
\end{subequations}

In the top right panel of Figure~\ref{fig:ZV_overview}, we compare the scaling of the different determinations of the local vector current renormalisation constant $Z_\mathrm{V}$.
The lattice spacings are extracted from a fit to the results of \cite{Bruno:2016plf}.
The deviation of the ratio of our two different determinations $Z_\mathrm{V}^\mathrm{k}$ and $Z_\mathrm{V}^\mathrm{f}$ from $1$ can be very well described by a term quadratic in $a$ with $\chi^2/\mathrm{d.o.f.}=0.91$, which confirms the expectation that the leading relative cutoff effects are dominantly of $\mathrm{O}(a^2)$.
The relative cutoff effects between our determination $Z_\mathrm{V}^\mathrm{f}$ and $Z_\mathrm{V}^\mathrm{G}$ from~\cite{Gerardin:2018kpy} are best described by a fit to $1+c\cdot a$ with $\chi^2/\mathrm{d.o.f.}=0.35$. We can also model the data with a fit to $1+c_1\cdot a^2+c_2\cdot a^3$ with $\chi^2/\mathrm{d.o.f.}=1.14$. A priori, a scaling behaviour with leading $\mathrm{O}(a)$ effects is unexpected, as both determinations are supposedly valid up to cutoff effects of $\mathrm{O}(a^2)$. For this reason we advocate the two-parameter fit, which is also depicted in Figure~\ref{fig:ZV_overview}, and conclude that our determination agrees with the one of Ref.~\cite{Gerardin:2018kpy} up to the expected ambiguities.

Moreover, the relative scaling between the result from the chirally rotated Schrödinger functional $Z_\mathrm{V}^\chi$ \cite{DallaBrida:2018tpn} and ours on $Z_\mathrm{V}^\mathrm{f}$ shows $\mathrm{O}(a^3)$ effects of significant size in addition to the leading $\mathrm{O}(a^2)$ ones, which was also observed in \cite{Gerardin:2018kpy} for their result. The deviation from $Z_\mathrm{V}^\chi/Z_\mathrm{V}^\mathrm{f}=1$ is described best by both a quadratic and a cubic term with $\chi^2/\mathrm{d.o.f.}=0.42$, while other two-parameter fits are incompatible. This is perfectly in line with the theoretical expectation of relative cutoff effects of at least $\mathrm{O}(a^2)$, but it may have practical implication on the subsequent determination of the vector current's improvement coefficient which turns out to depend rather sensitively on the choice of the renormalisation constant. To corroborate this point, we take a closer look at the ratio $(Z_\mathrm{A}/Z_\mathrm{V})^2$ (i.e., the relevant ratio entering the expression for $\cv$, as we will see in the next section) for different variants of $Z_\mathrm{V}$ in the bottom right panel of Figure~\ref{fig:ZV_overview}. While the ratio of the two determinations based on the same method seems to deviate from unity by effects dominated by $\mathrm{O}(a)$, its ``mixed'' ratios appear to exhibit additional ambiguities of $\mathrm{O}(a^2)$ that dominate for the coarser lattice spacings of interest.
\subsection{Check of the LCP condition} \label{subsec:lcp_violation}
Finally, we want to assess the systematic error that could be induced in our results by a violation of the LCP condition.
Before discussing its quantitative outcome, let us recall the idea behind
it: Imposing improvement and renormalisation conditions based on chiral
Ward identities implies chiral symmetry to be fully recovered only in the
continuum limit.
Consequently, the choice of these conditions matters at the cutoff level,
i.e., at a fixed value of the lattice spacing, the numerical results may
differ ---~sometimes even considerably~--- between any two such choices.
However, when adhering to the prescription of first defining an LCP
(by fixing all dimensionful parameters in terms of a physical scale) along
which the continuum limit is taken, these intrinsic differences should not
be interpreted as a systematic error.
Rather, since this LCP approach yields functions $c(g_0^2)$ and $Z(g_0^2)$
as the lattice spacing is varied and, accordingly, different functions for
another choice of the LCP, the difference between any two of these choices
will be, within errors, a smooth function of $g_0^2$ that vanishes
asymptotically at least $\propto a$ or $\propto a^2$ if $\rmO(a)$ improvement is
implemented.
In other words, following an LCP ensures that cutoff effects obey a
smooth functional dependence on $g_0^2$ and the particular choice of LCP becomes irrelevant in the continuum limit.
Hence, from this perspective, the relevant systematic error is determined
by the precision, to which a chosen LCP is realised and can be followed.
\begin{figure}[t]
	\centering
	\begin{subfigure}{.5\textwidth}
		\centering
		\includegraphics[width=\linewidth]{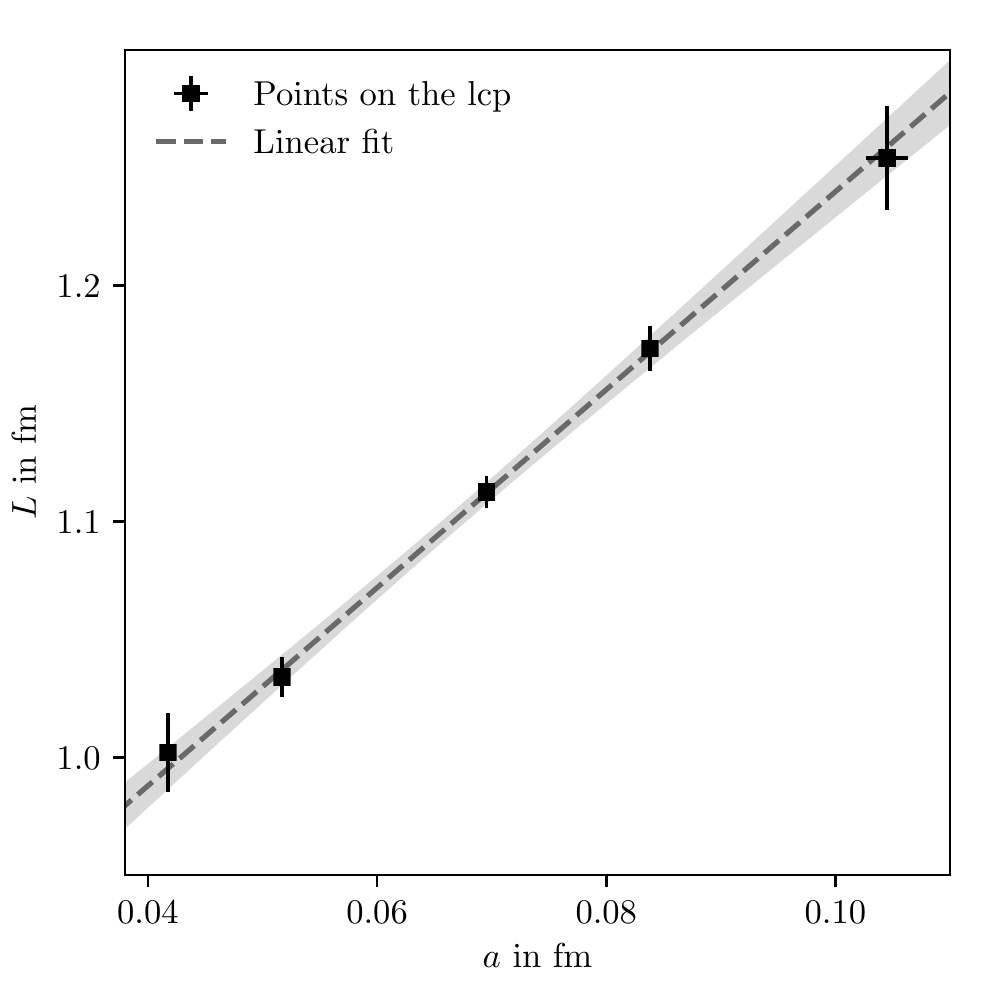}
	\end{subfigure}%
	\begin{subfigure}{.5\textwidth}
		\centering
		\includegraphics[width=\linewidth]{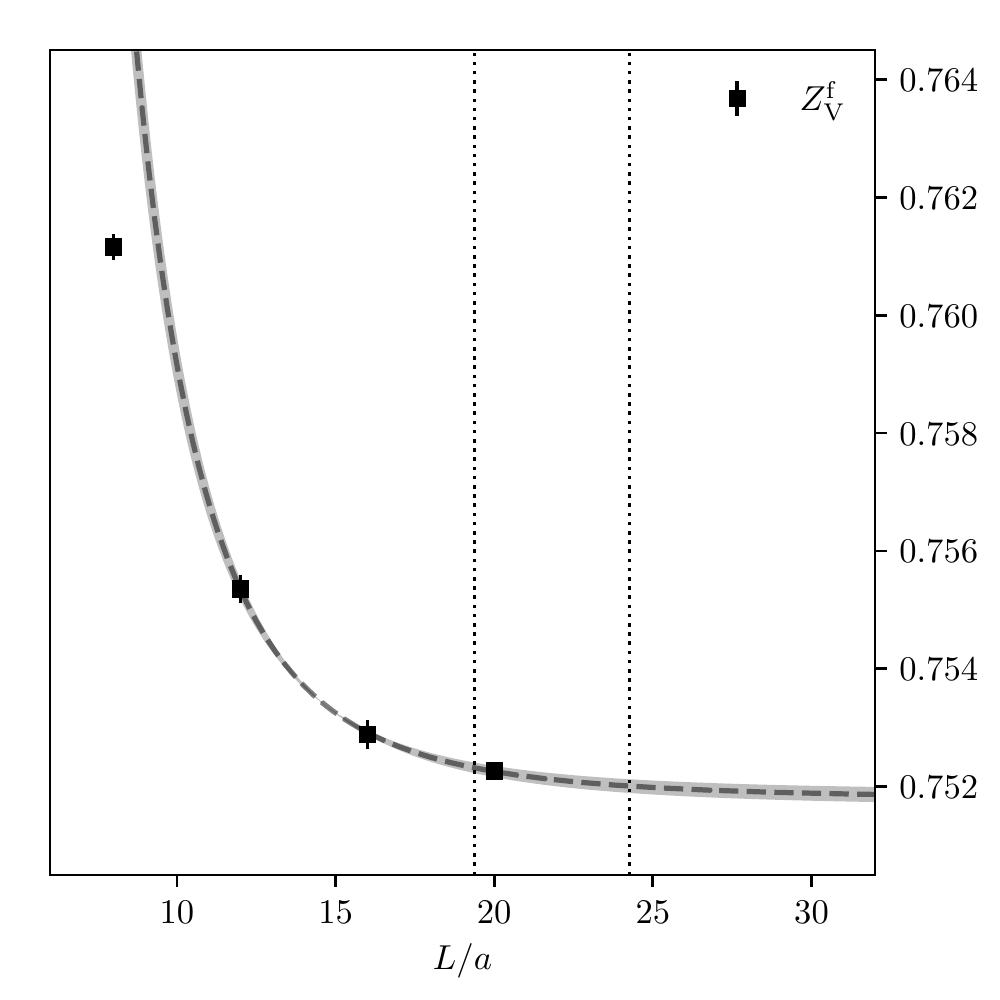}
	\end{subfigure}
	\caption{{\em Left:} Physical extent $L$ of the lattices on the LCP used for the renormalisation of the vector current (and also for $c_\mathrm{V}$ and $c_{\tilde{\mathrm{V}}}$) as a function of the lattice spacing $a$; the latter is deduced from the scale setting of Ref.~\cite{Bruno:2016plf}. {\em Right:} Volume dependence of $Z_\mathrm{V}^\mathrm{f}$ for $\beta=3.676$. The data can be well described by a fit linear in $(a/L)^4$ when excluding the smallest volume. The dotted vertical lines indicate the range of corresponding physical values of $L$ realized on our LCP.}
	\label{fig:finite_volume_effects}
\end{figure}

As the LCP used in this work was originally defined based on the perturbative two-loop $\beta$-function, we expect the deviation from the intended value of $L$ in physical units to be proportional to the lattice spacing $a$.
Information, on how accurately the chosen LCP condition is met for our
simulation parameters, can be gathered from the left panel of Figure~\ref{fig:finite_volume_effects}. The deviation can indeed be described by a function linear in the lattice spacing~$a$. Based on this result we expect any effects of this deviation on our quantities of interest to be of the next-to-leading order in $a$ (i.e., $\mathrm{O}(a^2)$ for the renormalisation of the vector current in the $\mathrm{O}(a)$ improved theory).

To get a feeling for the size of these effects we explicitly investigate the volume dependence of $Z_\mathrm{V}$ at a fixed value of $\beta=3.676$. For this purpose we have generated additional gauge ensembles, summarised in Table~\ref{tab:finite_volume_parameters}, which coincide with the ensembles labeled C in Table~\ref{tab:gauge_parameters} except for the lattice geometry. We evaluate $Z_\mathrm{V}^\mathrm{f}$ at the chiral point on these ensembles as described in the previous subsection. 
\begin{table}[t]
	\centering
	\renewcommand{\arraystretch}{1.25}
	\setlength{\tabcolsep}{3pt}
	\begin{tabular}{llrll}
		\toprule
		$L^3\times T/a^4$   & $\kappa$   & MDU     & $am$                     & $Z_\mathrm{V}^\mathrm{f}$   \\
		\midrule
		$16^3\times 24$     & 0.13719    & 15024 & $-$0.00095(14)           & 0.75205(20)               \\
		$16^3\times 24$     & 0.1369     & 12320 & \phantom{$-$}0.00939(21) & 0.76112(18)               \\
		&            &         & \phantom{$-$}0.0         & 0.75288(25)               \\
		\midrule
		$12^3\times 18$     & 0.13719    & 14304 & $-$0.00125(20)           & 0.75428(19)               \\
		$12^3\times 18$     & 0.1369     & 17392 & \phantom{$-$}0.00913(19) & 0.76323(20)               \\
		&            &         & \phantom{$-$}0.0         & 0.75535(24)               \\
		\midrule
		$8^3\times 12$      & 0.13719    & 32000 & $-$0.00153(19)           & 0.75979(20)               \\
		$8^3\times 12$      & 0.1369     & 32000 & \phantom{$-$}0.00882(19) & 0.76904(19)               \\
		&            &         & \phantom{$-$}0.0         & 0.76116(23)               \\
		\bottomrule
	\end{tabular}
	\caption{Summary of simulation parameters for the additional gauge configuration ensembles at $\beta=3.676$ used for the study of finite volume effects. The time extent of the ensembles is even in contrast to the ones listed in Table~\ref{tab:gauge_parameters}. This is due to the use of an updated version of \texttt{openQCD}. MDU denotes the total number of molecular dynamics units. The results at vanishing quark mass were obtained via the procedure described in the previous subsection.}
	\label{tab:finite_volume_parameters}
\end{table}
Our estimates on the impact of and possible systematic uncertainties due to
deviations from the chosen LCP are displayed in the right panel of Figure~\ref{fig:finite_volume_effects}. Apart from the the result at $L/a=8$, the data can be well described by a constant plus a term proportional to $(a/L)^4$. The vertical dashed lines indicate the span of volumes which are part of the LCP. The empirical finding is that volumes of intermediate extents $\sim 1.2\,$fm, employed in this study, correspond to a regime where effects owing to a not exactly fixed physical volume\footnote{Note that in the Schrödinger functional framework as employed here, the physical length scale $L$ is in one-to-one correspondence to the renormalisation scale of a properly defined finite-volume coupling, which in principle could have also been kept fixed for defining an LCP.}$\!\!$ are mild but still of a similar magnitude as the statistical uncertainties achieved here.
In the spirit of the previous line of reasoning, however, we decide to not include this uncertainty into our final results, because the deviations from the LCP are of $\mathrm{O}(a^2)$ or higher and therefore beyond the level that would have to be considered as a contribution to the systematic error.

\subsection{Mass dependent $\mathrm{O}(a)$ improvement}
For the sake of completeness we would like to briefly address the quark mass dependent pieces in the renormalisation patterns of the vector current according to \eqs{eq:def_ZVf} and (\ref{eq:def_ZVf}).
To achieve full $\mathrm{O}(a)$ improvement of the local current at finite quark mass, one not only requires (besides $c_\mathrm{V}$) the renormalisation constant $Z_\mathrm{V}$ in the chiral limit, but also the coefficients $\bar{b}_\mathrm{V}$ and $b_\mathrm{V}$ which parametrise the sea and valence quark mass dependence, respectively. Both coefficients are already non-perturbatively known for our choice of action from Refs.~\cite{Fritzsch:2018zym,Gerardin:2018kpy}. Moreover, as the mass range width of our gauge ensembles is not ideal for a detailed study of these mass dependent effects for all $g_0^2$, we restrict ourselves to an accurate determination of $b_\mathrm{V}$ at only one value of the bare coupling, corresponding to $\beta=3.414$, which we compare to the previous results.
The coefficient $b_\mathrm{V}$ is readily accessible through evaluating the logarithm of the renormalisation condition for $\zv$ at different valence and vanishing sea quark masses. The slope with respect to the bare quark mass is then proportional to the $b$-coefficient of interest. Thereby, from the above renormalisation conditions (\ref{eq:def_ZVf}) and (\ref{eq:def_ZVk}), one arrives at two independent estimates of $b_\mathrm{V}$ in terms of ratios of correlation functions by mapping out the valence quark mass dependence at vanishing sea quark mass:
\begin{align}
b_\mathrm{V}^\mathrm{f} &= \frac{\partial }{\partial (am_\mathrm{q})}\log\bigg (\frac{F_1}{F_{\mathrm{V}}(x_0)}\bigg)\bigg|_{m_\mathrm{q}=0} +\mathrm{O}(a)\,,\quad \text{at } \mathrm{Tr}[aM_\mathrm{q}]=0 \,,
\label{eq:def_bVf}\\
b_\mathrm{V}^\mathrm{k} &= \frac{\partial }{\partial (am_\mathrm{q})}\log\bigg (\frac{K_1}{K_{\mathrm{V}}(x_0)}\bigg)\bigg|_{m_\mathrm{q}=0} +\mathrm{O}(a)\,,\quad \text{at } \mathrm{Tr}[aM_\mathrm{q}]=0 \,.
\label{eq:def_bVk}
\end{align}

\begin{figure}[t]
	\centering
	\begin{subfigure}{.5\textwidth}
		\centering
		\includegraphics[width=\linewidth]{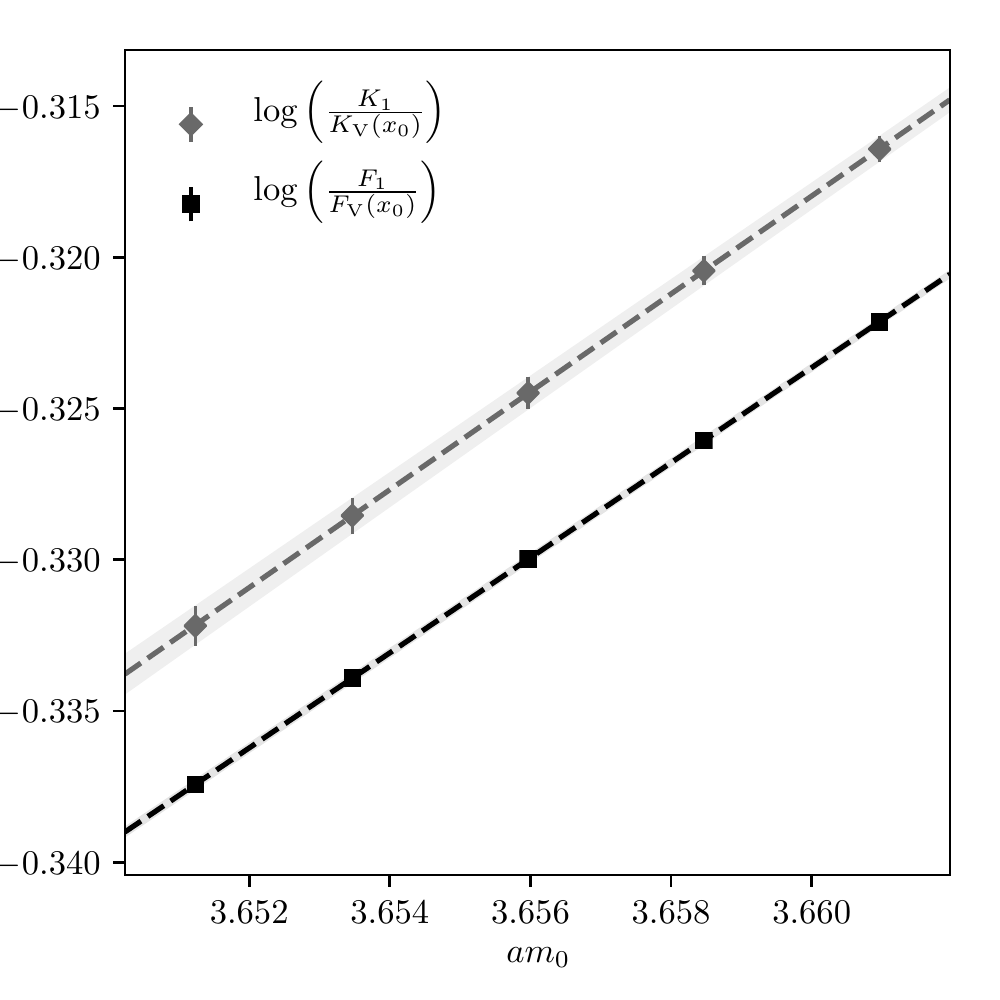}
	\end{subfigure}%
	\begin{subfigure}{.5\textwidth}
		\centering
		\includegraphics[width=\linewidth]{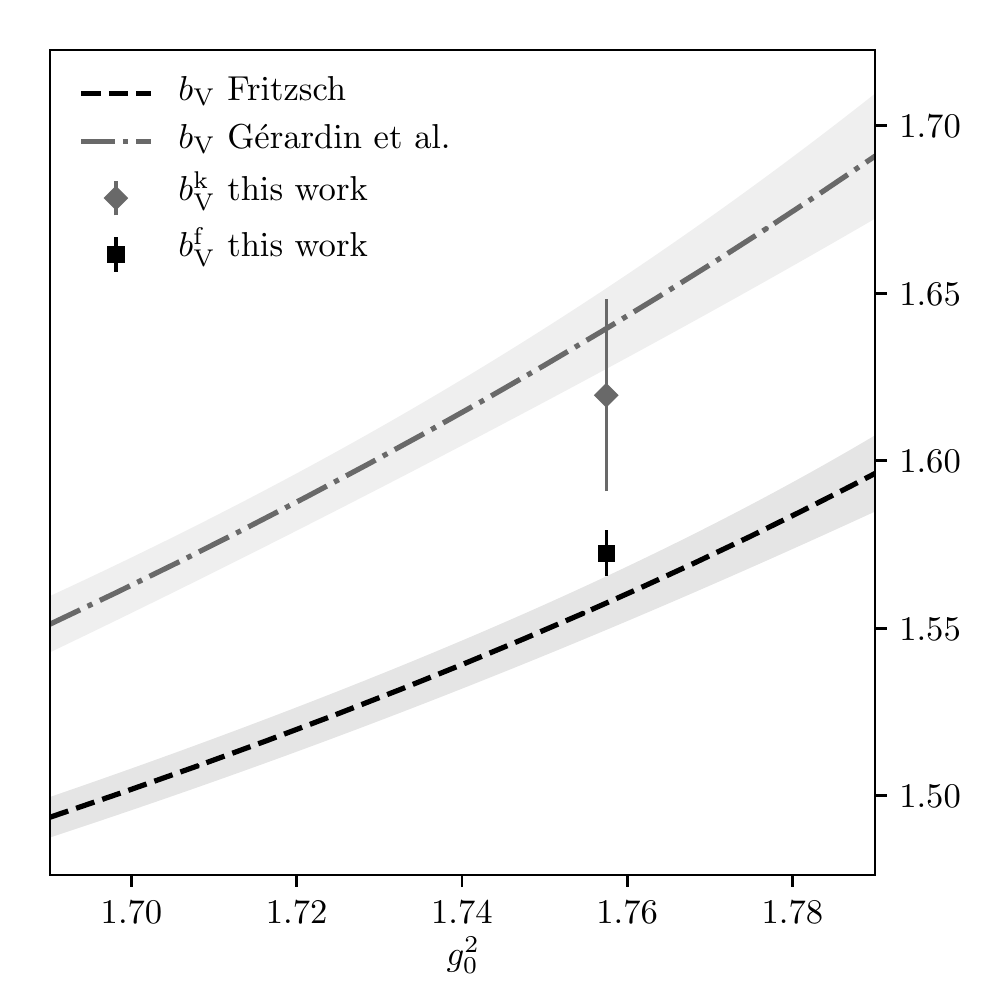}
	\end{subfigure}
	\caption{{\em Left:} Valence quark mass dependence of the log-terms in \eqs{eq:def_bVf} and (\ref{eq:def_bVk}) at $\beta=3.414$ (ensemble E1k2). For both variants, the dashed lines and the surrounding shaded areas describe a linear fit to the data points, whose slope yields $b_\mathrm{V}^{\{\mathrm{f},\mathrm{k}\}}$. {\em Right:} Comparison of our two determinations of $b_\mathrm{V}$ at $\beta=3.414$ with the results of \cite{Fritzsch:2018zym,Gerardin:2018kpy}, where the latter are represented as functions of the bare coupling squared.}
	\label{fig:bV_results}
\end{figure}

For the numerics we choose ensemble E1k2 ($\beta=3.414$), which fulfills the condition of vanishing sea quark masses within statistical precision, and calculate the relevant ratios of correlation functions at five different, equally spaced valence quark masses, as shown in the left part of Figure~\ref{fig:bV_results}.
In the right part of Figure~\ref{fig:bV_results} we compare our two determinations with the results from the literature. Our $b_\mathrm{V}^\mathrm{f}$ is very closely consistent with the results of \cite{Fritzsch:2018zym}, which emerge from an almost identical computational setup but employ a smaller physical volume. This supports the claim made in \cite{Fritzsch:2018zym} that the volume dependence of the $b$-coefficients is not statistically significant. Although our result from the other estimator, $b_\mathrm{V}^\mathrm{k}$, has a considerably larger statistical uncertainty, it is perfectly compatible with the result of \cite{Gerardin:2018kpy}, in particular when keeping in mind that intrinsic $\mathrm{O}(a)$ ambiguities among the $b$-coefficients from different improvement (and realisations of) constant physics conditions must be expected in any case.

\section{$\mathrm{O}(a)$ improvement of the vector current}
\label{sec:improvement}
The method for the determination of the vector current improvement coefficient based on chiral Ward identities has first appeared and been tested in \cite{Guagnelli:1997db} in the quenched case. Recently, it has also been applied to QCD with $N_\mathrm{f}=2+1$ dynamical fermions~\cite{Gerardin:2018kpy}, using the same $\rmO(a)$ improved lattice action as we do, but implementing it directly on large-volume ensembles generated by the CLS effort. In this section we will briefly recapitulate the derivation of the improvement condition within our intermediate-volume approach realised by the Schrödinger functional setup and then propose a slight variation that helps to tame the large higher-order ambiguities in $(Z_\mathrm{A}/Z_\mathrm{V})^2$, anticipated in the previous section, which now become influential.
\subsection{Improvement condition}
We now consider infinitesimal axial vector variations of the quark fields
\begin{align}
\delta_\mathrm{A}^a\bar\psi(x)\approx\mathrm{i}\bar\psi(x)T^a\gamma_5\,,\quad 
\delta_\mathrm{A}^a\psi(x)\approx\mathrm{i}T^a\gamma_5\psi(x)\,.
\end{align}
The improvement condition studied in the following is based on the invariance of the expectation value of composite operators under these rotations
\begin{align}
	\big\langle \mathit{O}\,\delta_\mathrm{A}S\big\rangle = \big\langle \delta_\mathrm{A}\mathit{O}\big\rangle\,,
\end{align}
which is basis for the axial vector Ward identity presented here in its integrated form
\begin{align}
	\begin{split}
		&\int \mathrm{d}^3\mathbf{x} \,\big\langle [A_0^a(t_2,\mathbf{x})-A_0^a(t_1,\mathbf{x})] \mathit{O}^b(y)\mathit{O}^{\prime c}(z)\big\rangle \\
		&- 2m\int \mathrm{d}^3\mathbf{x}\int_{t_1}^{t_2}\mathrm{d}x_0 \,\big\langle P^a(x_0,\mathbf{x}) \mathit{O}^b(y)\mathit{O}^{\prime c}(z)\big\rangle \\
		=& -\,\big\langle  [\delta_\mathrm{A}^a\mathit{O}^{b}(y)]\mathit{O}^{\prime c}(z) \big\rangle\,,
	\end{split}
\end{align}
where we restrict the axial variation to the space-time region between $t_1$ and $t_2$ and impose $z_0<t_1<y_0<t_2$.
When we identify $\mathit{O}^{\prime c}$ with $\mathcal{Q}_k^c$ and $\mathit{O}^b$ with $A_k^b$, which changes flavour as well as Dirac structure under small axial vector rotations
\begin{align}
\delta_\mathrm{A}^aA_k^b(x)=-\mathrm{i}\epsilon^{abc}V_k^c(x)\,,
\end{align}
we arrive at
\begin{align}
\label{eq:cv_ward_identity}
	\begin{split}
		&Z_\mathrm{A}^2\Big[k_\mathrm{AA}^\mathrm{I}(t_2,x_0)-k_\mathrm{AA}^\mathrm{I}(t_1,x_0)-2m\tilde{k}_\mathrm{PA}(t_1,t_2,x_0)  \Big]\\
		=\,&Z_\mathrm{V}\Big[k_\mathrm{V}(x_0)+ac_\mathrm{V}\tilde{\partial}_0k_\mathrm{T}(x_0)\Big]+ \mathrm{O}(a^2)\,.
	\end{split}
\end{align}
The finite renormalisation constants $Z_\mathrm{A}$ and $Z_\mathrm{V}$ and the $\mathrm{O}(a)$ improvement terms proportional to $c_\mathrm{A}$ and $c_\mathrm{V}$, respectively, assure that the Ward identity is valid up to quadratic corrections in the lattice spacing. (Explicit expressions of the Schrödinger functional boundary-to-bulk correlation functions involved are summarised in Appendix~\ref{sec:sf_correlation_functions}.)

This Ward identity can be rearranged to yield a first improvement condition for $c_\mathrm{V}$, viz.,
\begin{align}
	\begin{split}
		&c_\mathrm{V}(x_0,t_1,t_2;Z_\mathrm{A}, Z_\mathrm{V}, c_\mathrm{A})\\
		=\,& \frac{Z_\mathrm{A}^2}{Z_\mathrm{V}}\frac{\big[k_{\mathrm{A}\mathrm{A}}^\mathrm{I}(t_2, x_0) - k_{\mathrm{A}\mathrm{A}}^\mathrm{I}(t_1, x_0)-2m\tilde{k}_{\mathrm{P}\mathrm{A}}(t_1, t_2, x_0)\big]}{a\partial_0k_\mathrm{T}(x_0)} - \frac{k_\mathrm{V}(x_0)}{a\partial_0k_\mathrm{T}(x_0)} + \mathrm{O}(a)\,.
        \end{split}
\label{eq:imprcond_cV_1}
\end{align}
For the moment we state the functional (resp.\ parametric) dependence on the Euclidean times $x_0$, $t_1$ and $t_2$ as well as on the renormalisation and improvement factors $Z_\mathrm{A}$, $Z_\mathrm{V}$ and $c_\mathrm{A}$ explicitly. The superscript `$\mathrm{I}$' labels improved correlation functions as detailed in Appendix~\ref{sec:sf_correlation_functions}. In this improvement condition, $c_\mathrm{V}$ arises as the difference of two potentially large terms, where only one of the two is sensitive to a change in $Z_\mathrm{A}$. Therefore, even a sub-permille uncertainty on $Z_\mathrm{A}$ can easily propagate into a relative error well above five percent on the final error and may hence become the dominant contribution to the overall uncertainty on $c_\mathrm{V}$ at the end.

In order to suppress the relative higher-order effects observed in the determination \cite{DallaBrida:2018tpn} of the normalisation constants $Z_\mathrm{A}$ and $Z_\mathrm{V}$ via the chirally rotated Schrödinger functional, we propose an alternative improvement condition which exploits the previously discussed vector Ward identity
\begin{align}
	\frac{F_1}{Z_\mathrm{V}F_\mathrm{V}(x_0)}=1+\mathrm{O}(a^2)
\label{eq:vectorWI}
\end{align}
to expand the fraction in the first term on the r.h.s.\ of eq.~(\ref{eq:imprcond_cV_1}). This gives another estimator for the improvement coefficient $c_\mathrm{V}$, labeled by $c_\mathrm{V, \mathrm{alt}}$,
\begin{align}
	\begin{split}
		&c_{\mathrm{V}, \mathrm{alt}}(x_0,t_1,t_2;Z_\mathrm{A}, Z_\mathrm{V}, c_\mathrm{A})\\
		=\,& \frac{Z_\mathrm{A}^2}{Z_\mathrm{V}^2}\frac{\big[k_{\mathrm{A}\mathrm{A}}^\mathrm{I}(t_2, x_0) - k_{\mathrm{A}\mathrm{A}}^\mathrm{I}(t_1, x_0)-2m\tilde{k}_{\mathrm{P}\mathrm{A}}(t_1, t_2, x_0)\big]}{a\partial_0k_\mathrm{T}(x_0)}\frac{F_1}{F_\mathrm{V}(x_0)} - \frac{k_\mathrm{V}(x_0)}{a\partial_0k_\mathrm{T}(x_0)} + \mathrm{O}(a)\,,
	\end{split}
\label{eq:imprcond_cV_2}
\end{align}
and defines our alternative improvement condition under consideration. The expressions in eqs.~(\ref{eq:imprcond_cV_1}) and (\ref{eq:imprcond_cV_2}) are equivalent up to lattice spacing ambiguities of $\mathrm{O}(a)$ or higher, which corresponds to effects of $\mathrm{O}(a^2)$ or beyond in any matrix elements of the $\mathrm{O}(a)$ improved vector current. However, eq.~(\ref{eq:imprcond_cV_2}) benefits from the fact that the ratio $Z_\mathrm{A}^2/Z_\mathrm{V}^2$, being directly deducible from Ref.~\cite{DallaBrida:2018tpn}, has balanced powers of the $Z$-factors (in contrast to $Z_\mathrm{A}^2/Z_\mathrm{V}$ entering (\ref{eq:imprcond_cV_1})), from which one may expect an approximate cancellation of the dominating higher-order ambiguities as suggested by the dashed line in the bottom part of Figure~\ref{fig:ZV_overview} and thus also a reduced impact of these systematics on our result.

For the improvement coefficient of the point-split vector current, the two variants of the improvement condition look slightly different. For the original improvement condition we simply substitute $k_\mathrm{V}$ by $k_{\tilde{\mathrm{V}}}$ and use the fact that $Z_{\tilde{\mathrm{V}}}=1$:
\begin{align}
\label{eq:standard_cv_tilde}
\begin{split}
&c_{\tilde{\mathrm{V}}}(x_0,t_1,t_2;Z_\mathrm{A}, c_\mathrm{A})\\
=\,& Z_\mathrm{A}^2\frac{\big[k_{\mathrm{A}\mathrm{A}}^\mathrm{I}(t_2, x_0) - k_{\mathrm{A}\mathrm{A}}^\mathrm{I}(t_1, x_0)-2m\tilde{k}_{\mathrm{P}\mathrm{A}}(t_1, t_2, x_0)\big]}{a\partial_0k_\mathrm{T}(x_0)} - \frac{k_{\tilde{\mathrm{V}}}(x_0)}{a\partial_0k_\mathrm{T}(x_0)} + \mathrm{O}(a)\,.
\end{split}
\end{align}
To arrive at an alternative improvement condition, where in a similar spirit as before the ratio $Z_\mathrm{A}^2/Z_\mathrm{V}^2$ enters, we need to multiply the vector Ward identity (\ref{eq:vectorWI}) to the first term on the r.h.s.\ twice. This yields:
\begin{align}
	\begin{split}
		&c_{\tilde{\mathrm{V}}, \mathrm{alt}}(x_0,t_1,t_2;Z_\mathrm{A}, Z_\mathrm{V}, c_\mathrm{A})\\
		=\,& \frac{Z_\mathrm{A}^2}{Z_\mathrm{V}^2}\frac{\big[k_{\mathrm{A}\mathrm{A}}^\mathrm{I}(t_2, x_0) - k_{\mathrm{A}\mathrm{A}}^\mathrm{I}(t_1, x_0)-2m\tilde{k}_{\mathrm{P}\mathrm{A}}(t_1, t_2, x_0)\big]}{a\partial_0k_\mathrm{T}(x_0)}\Bigg(\frac{F_1}{F_\mathrm{V}(x_0)}\Bigg)^2 - \frac{k_{\tilde{\mathrm{V}}}(x_0)}{a\partial_0k_\mathrm{T}(x_0)} + \mathrm{O}(a)\,.
	\end{split}
\end{align}
\subsection{Analysis details}
\begin{figure}[t]
	\centering
	\begin{subfigure}{.5\textwidth}
		\centering
		\includegraphics[width=\linewidth]{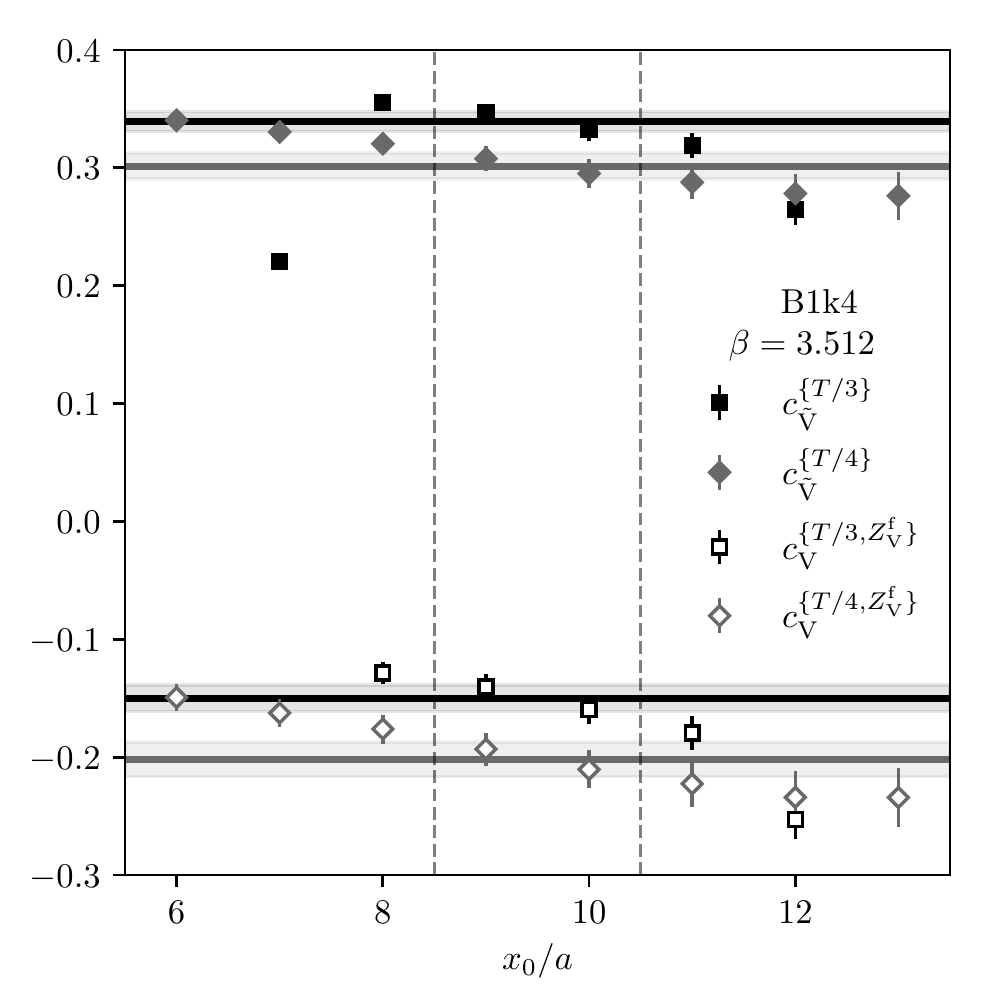}
	\end{subfigure}%
	\begin{subfigure}{.5\textwidth}
		\centering
		\includegraphics[width=\linewidth]{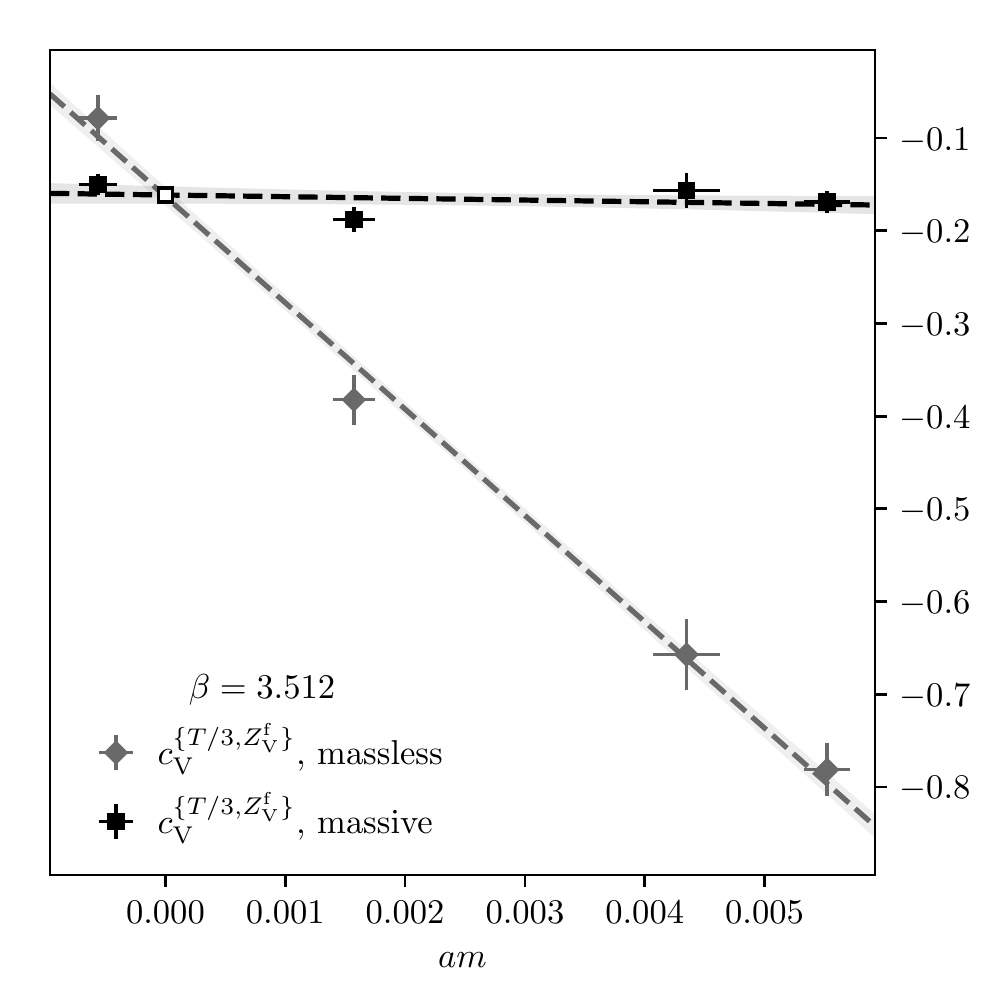}
	\end{subfigure}
	\caption{{\em Left:} Time dependence of different variants of the improvement coefficient for the conserved and the local vector current on ensemble B1k4 ($\beta=3.512$). The dashed vertical lines enclose the plateau region, while the horizontal lines with shaded regions correspond to the plateau values and their uncertainties. {\em Right:} Chiral extrapolation of $c_\mathrm{V}^{\lbrace T/3, Z_\mathrm{V}^\mathrm{f} \rbrace}$ for $\beta=3.512$ with and without the mass term in the Ward identity. The dashed lines correspond to linear fits to the data, and the shaded regions indicate the uncertainties of the fits. The open symbol marks the value in the chiral limit.}
	\label{fig:cV_extraction}
\end{figure}
We evaluate the improvement conditions for the local and the conserved (i.e., point-split) vector current with two different operator positions $t_1=T-t_2 \in \lbrace T/4,T/3 \rbrace$ and average over the two central values of $x_0$. For the axial vector improvement coefficient $c_\mathrm{A}$ we employ the result of \cite{Bulava:2015bxa}, while for the vector current normalisation constant we try both, the results $Z_\mathrm{V}^\mathrm{f}$ from our determination reported here and the values $Z_\mathrm{V}^\chi$ obtained within the chirally rotated Schrödinger functional setup \cite{DallaBrida:2018tpn}. For the axial vector current renormalisation constant we only rely on the result of \cite{DallaBrida:2018tpn} labeled $Z_\mathrm{A}^\chi$, because the accuracy reached for this quantity in the standard Schrödinger functional setup \cite{Bulava:2016ktf} is not sufficient for a satisfactory precision of the vector current improvement constants and would cap the final absolute errors above $\approx0.1$. Hence, from hereon we simplify the notation specifying our particular choices to define the two improvement coefficients for each of the two vector currents to
\begin{align}
	\frac{1}{2}\sum_{x_0=T/2-1}^{T/2+1}c_{X,(\mathrm{alt})}(x_0,t_1,T-t_1;Z_\mathrm{A}^\chi, Z_\mathrm{V}, c_\mathrm{A}) = c_{X,(\mathrm{alt})}^{\lbrace t_1, Z_\mathrm{V} \rbrace}\,,\quad X=\mathrm{V},\tilde{\mathrm{V}}\,.
\label{eq:cV_short}
\end{align}
Note that in the case of the standard improvement condition for the conserved (i.e., point-split) current, eq.~(\ref{eq:standard_cv_tilde}), no $Z_\mathrm{V}$-dependence arises.

$x_0$-dependences of the improvement conditions on an exemplary ensemble are presented in the left part of Figure~\ref{fig:cV_extraction}. The fact that the local estimator does not develop a clear plateau but rather slightly descends towards larger values of $x_0$ was also seen in the similar study \cite{Gerardin:2018kpy} as well as in the computation of the tensor current improvement constant \cite{Chimirri:2019xsv}. We attribute this to the smaller Euclidean time distances of the operators contributing to the Ward identities in comparison to the PCAC relation that usually exhibits more distinct plateaux. Note, however, that by virtue of basing the extraction of improvement coefficients (and normalisation constants) on Ward identities formulated on the operator level and combining it with the constant physics idea, there is a natural freedom in choosing $x_0$ and $t_1$ when actually evaluating the given improvement condition in a numerical calculation, as long as all physical scales are kept fixed among the different ensembles. Therefore, even if the $x_0$-dependence of the estimators does not exhibit a plateau in the central region, taking an average over central values of $x_0$ as implied by eq.~(\ref{eq:cV_short}) is perfectly legitimate. In case of the operator position $t_1=T/3$, the plateau-like behaviour spans a shorter time separation as expected. For the reason of larger separation from the boundary and because of the slightly smaller statistical uncertainties, we prefer $t_1=T/3$ over $t_1=T/4$ and thus regard this choice as part of the definition of our estimators for $c_\mathrm{V}$; the latter choice (i.e., $t_1=T/4$) is only used for scaling analyses. In the right part of Figure~\ref{fig:cV_extraction} the chiral extrapolation for one specific improvement condition, $c_\mathrm{V}^{\lbrace T/3,Z_\mathrm{V}^\mathrm{f}\rbrace}$, is displayed. We compare the extrapolation with and without the explicit mass term (see, e.g., eq.~(\ref{eq:cv_ward_identity})) to illustrate that accounting for this term in the evaluation of the Ward identity leads to an almost flat quark mass dependence and a thereby very stable extrapolation. The remaining slope could in principle be compensated by a factor involving mass dependent pieces proportional to $b_\mathrm{A}$, $\bar{b}_\mathrm{A}$, $b_\mathrm{V}$ and $\bar{b}_\mathrm{V}$. As a sufficiently precise non-perturbative determination of the mass dependent improvement coefficients for the axial vector current is not available for our setup, we decide to omit this factor which has no effect on the result in the chiral limit. The results for the individual ensembles and chiral extrapolations are summarised in Tables~\ref{tab:cV_results} and \ref{tab:cV_tilde_results} in Appendix~\ref{app:tables}.
\begin{figure}[t]
	\centering
	\begin{subfigure}{.5\textwidth}
		\centering
		\includegraphics[width=\linewidth]{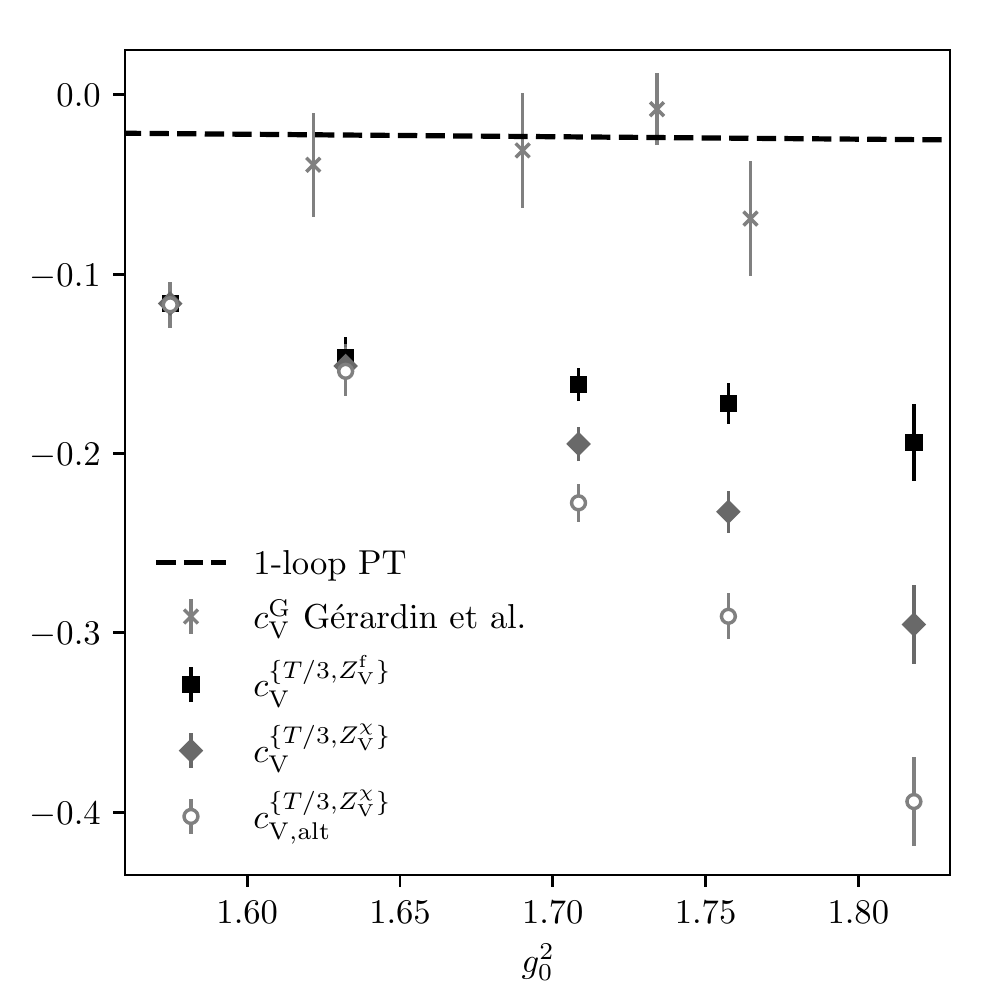}
	\end{subfigure}%
	\begin{subfigure}{.5\textwidth}
		\centering
		\includegraphics[width=\linewidth]{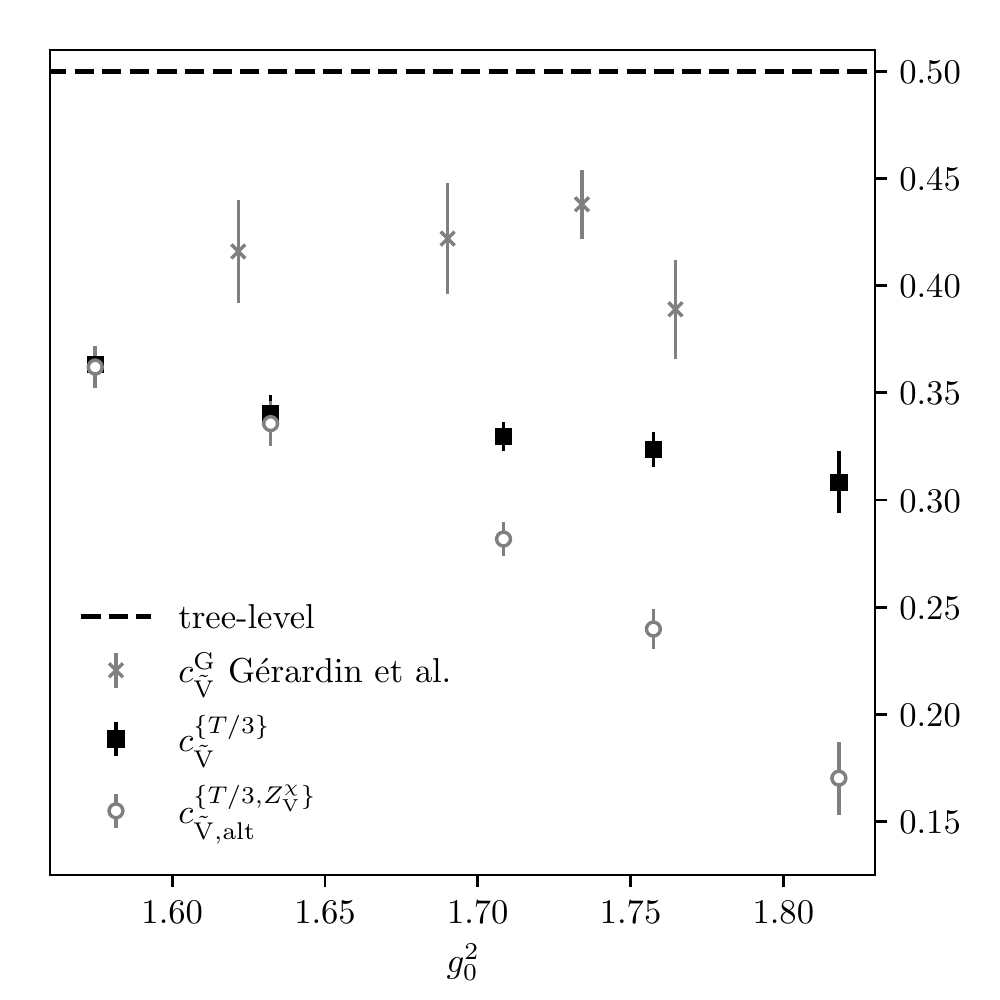}
	\end{subfigure}
	\caption{{\em Left:} Dependence of different determinations of $c_\mathrm{V}$ on the squared bare coupling in comparison to the results of \cite{Gerardin:2018kpy} and one-loop perturbation theory \cite{Taniguchi:1998pf}. {\em Right:} $g_0^2$-dependence of different determinations of $c_{\tilde{\mathrm{V}}}$ in comparison to the results of \cite{Gerardin:2018kpy} and the tree-level value.}
	\label{fig:cV_overview}
\end{figure}

In the left part of Figure~\ref{fig:cV_overview} we compare three different determinations of $c_\mathrm{V}$ from this work with the results of G\'{e}rardin et al.~\cite{Gerardin:2018kpy} and the one-loop prediction of \cite{Taniguchi:1998pf}. While the results obtained by G\'{e}rardin et al.\ are compatible with perturbation theory, all three of our determinations strongly deviate from this and only come close to the expected one-loop prediction for $c_\mathrm{V}$ as $g_0\to 0$. In particular, these three different variants agree within errors for the two smallest lattice spacings and show larger (albeit monotonic) spreads for the coarser ones. The two determinations $c_\mathrm{V}^{\lbrace T/3, Z_\mathrm{V}^\mathrm{f} \rbrace}$ and $c_\mathrm{V}^{\lbrace T/3, Z_\mathrm{V}^\chi \rbrace}$, which are sensitive to the higher-order ambiguities in $Z_\mathrm{A}^\chi$, show much milder scaling with $g_0^2$ compared to $c_{\mathrm{V}, \mathrm{alt}}^{\lbrace T/3, Z_\mathrm{V}^\chi \rbrace}$ with balanced powers of $Z$-factors, where the leading higher-order effects are supposed to be almost canceled. This suggests that these higher-order effects counteract the leading $\mathrm{O}(a)$ effects and imply the observed flat, somewhat non-uniform scaling behaviour with $g_0^2$, both present in the result of \cite{Gerardin:2018kpy} and our determination $c_\mathrm{V}^{\lbrace T/3, Z_\mathrm{V}^\mathrm{f} \rbrace}$. 
For these reasons we finally decide to advocate $c_{\mathrm{V}, \mathrm{alt}}^{\lbrace T/3, Z_\mathrm{V}^\chi \rbrace}$ as our preferred determination and expect it to be least affected by ambiguities in the ratio $Z_\mathrm{A}/Z_\mathrm{V}$.

A similar observation can also be made for the $g_0^2$-dependence of the improvement coefficient, $c_{\tilde{\mathrm{V}}}$, of the conserved vector current displayed in the right part of Figure~\ref{fig:cV_overview}. Since the standard improvement condition does not depend on $Z_\mathrm{V}$, we display only two variants here. As in the case of the local current, our results deviate much more from the tree-level formula in comparison to the result of \cite{Gerardin:2018kpy}. Again we observe a rather mild and less smooth dependence on the (squared) bare coupling for the result of G\'{e}rardin et al.\ and our determination, $c_{\tilde{\mathrm{V}}}^{\lbrace T/3\rbrace}$, which we attribute to the influence of higher-order effects in $Z_\mathrm{A}^\chi$. Therefore, also in this case we decide to prefer our alternative determination $c_{\tilde{\mathrm{V}}, \mathrm{alt}}^{\lbrace T/3,Z_\mathrm{V}^\chi\rbrace}$ over the former, for which we can well assume the dominant higher-order ambiguities to largely cancel.
\subsection{Final results}
\label{subsec:improvement_res}
\begin{figure}[t!]
	\centering
	\begin{subfigure}{.5\textwidth}
		\centering
		\includegraphics[width=\linewidth]{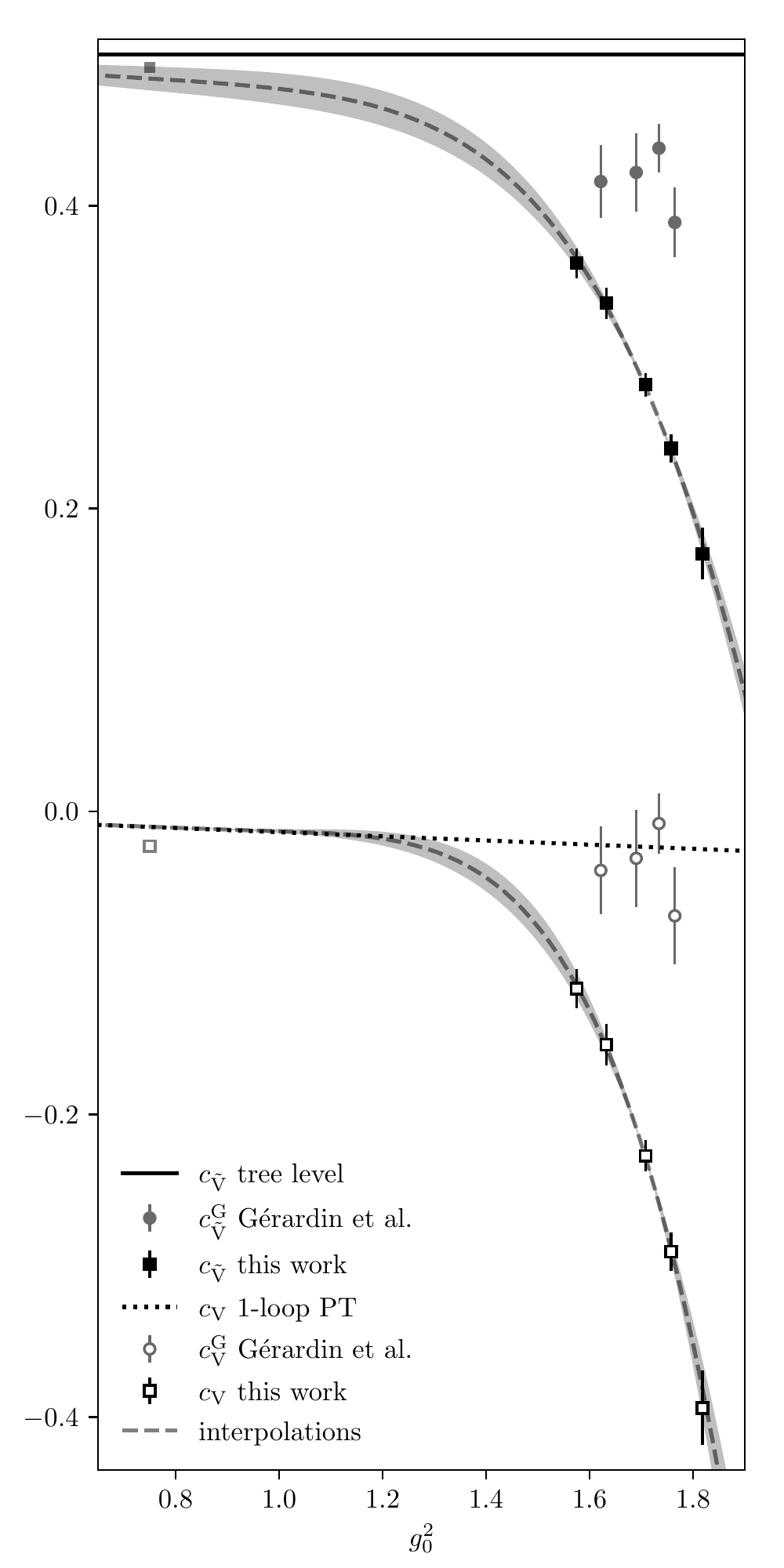}
	\end{subfigure}%
	\begin{subfigure}{.5\textwidth}
		\centering
		\includegraphics[width=0.997\linewidth]{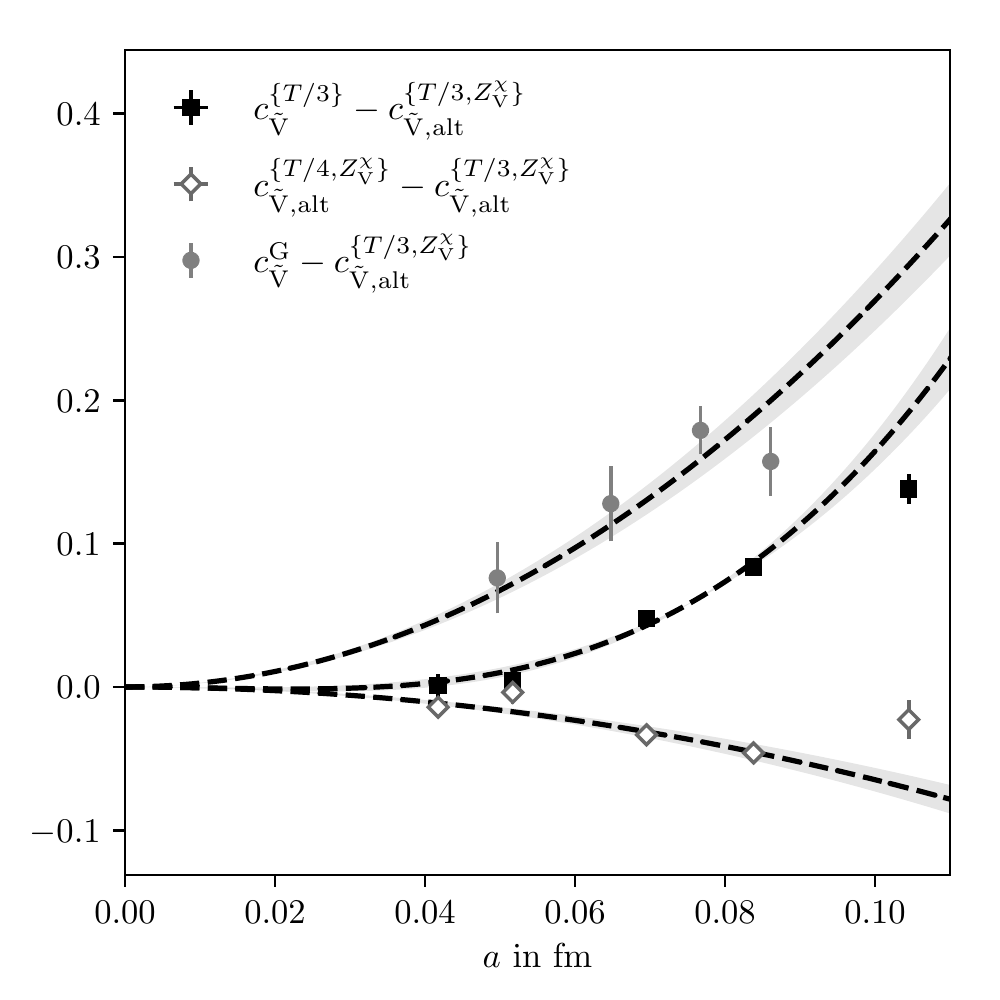}
		\includegraphics[width=0.997\linewidth]{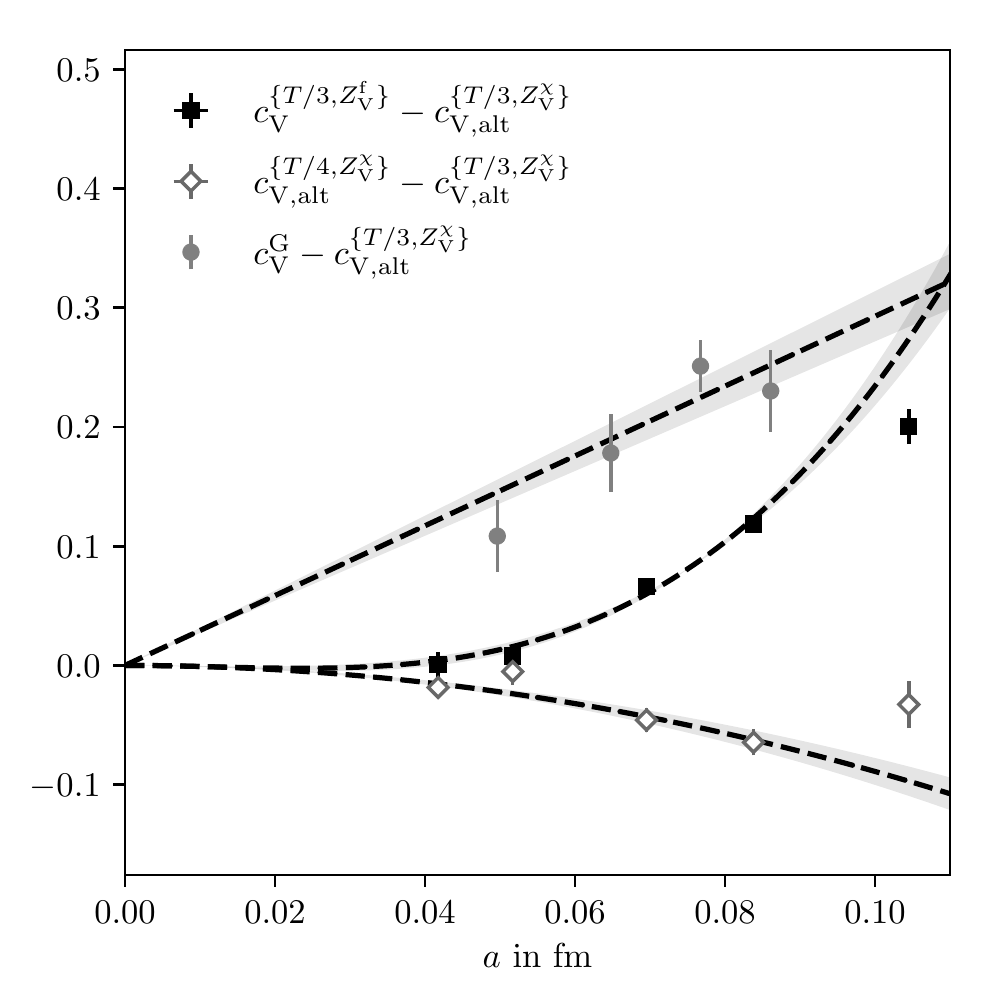}
	\end{subfigure}
	\caption{{\em Left:} Preferred non-perturbative determinations of $c_{\tilde{\mathrm{V}}}$ and $c_\mathrm{V}$ as functions of $g_0^2$ in the region, where they differ from the perturbative predictions, in comparison to the results of G\'{e}rardin et al. \cite{Gerardin:2018kpy} labeled by the superscript `$\mathrm{G}$'. The shaded points at $g_0^2=0.75$, i.e., lying deeply in the perturbative domain, are not included in the fits, as they do not lie on the LCP, but provide evidence that the improvement conditions approach the perturbative results towards the continuum limit as expected. {\em Top right:} Continuum limit of the difference of various determinations of $c_{\tilde{\mathrm{V}}}$ with our preferred variant. {\em Bottom right:} Continuum limit of the difference of various determinations of $c_{\mathrm{V}}$ with our preferred variant. In both figures on the right, the dashed lines describe the functional forms that describe the lattice spacing dependence best, as detailed in the text.}
	\label{fig:cV_interpolations}
\end{figure}
The left part of Figure~\ref{fig:cV_interpolations} collects our preferred determinations of the improvement coefficients of the local and conserved vector currents, respectively, in comparison to the perturbative predictions and the results of \cite{Gerardin:2018kpy}. For brevity we refer to these as $c_\mathrm{V}$ and $c_{\tilde{\mathrm{
V}}}$. The figure also displays the corresponding continuous interpolations of these final non-perturbative results in terms of $g_0^2$. The formulae are motivated by the leading term in the perturbative relation between the lattice spacing and the $\beta$-function, with universal coefficient $b_0=9/(4\pi)^2$ for $N_\mathrm{f}=3$, constrained by the one-loop~\cite{Taniguchi:1998pf} (tree-level) perturbative prediction for $g_0^2\rightarrow 0$ in case of $c_\mathrm{V}$ ($c_{\tilde{\mathrm{V}}}$). The $g_0^2$-dependence is then best described by a parametrisation of the following form:
\begin{subequations}
\label{eq:interpolation_cV}
\begin{align}
c_\mathrm{V}(g_0^2)&=-0.01030C_\mathrm{F}g_0^2\,\Big[1+\exp\big\lbrace-1/(2b_0g_0^2)\big\rbrace\big(c_\mathrm{V}^{(0)}+c_\mathrm{V}^{(1)}g_0^2\big)\Big]\,,
\\
c_{\tilde{\mathrm{V}}}(g_0^2)&=0.5-c_{\tilde{\mathrm{V}}}^{(0)}g_0^2\,\Big[1+\exp\big\lbrace-1/(2b_0g_0^2)\big\rbrace c_{\tilde{\mathrm{V}}}^{(1)}g_0^2\Big]\,,   
\end{align}
with
\begin{align}
c_\mathrm{V}^{(i)} &= 
\begin{pmatrix}
\begin{tabular}{@{}*{1}{S[table-format = +2.3e+1]}}
-3.039e+03 \\
+2.649e+03
\end{tabular}
\end{pmatrix}\,,
\quad \operatorname{cov}(c_\mathrm{V}^{(i)})=
\begin{pmatrix}
\begin{tabular}{@{}*{2}{S[table-format = +3.5e+1]}}
+1.97358e+06 & -1.15750e+06 \\
-1.15750e+06 & +6.78869e+05
\end{tabular}
\end{pmatrix}\,,\\
c_{\tilde{\mathrm{V}}}^{(i)} &= 
\begin{pmatrix}
\begin{tabular}{@{}*{1}{S[table-format = +2.3e+1]}}
+2.112e-02 \\
+5.065e+02
\end{tabular}
\end{pmatrix}\,,
\quad \operatorname{cov}(c_{\tilde{\mathrm{V}}}^{(i)})=
\begin{pmatrix}
\begin{tabular}{@{}*{2}{S[table-format = +3.5e+1]}}
+1.10461e-04 & -3.15408e+00 \\
-3.15408e+00 & +9.00608e+04
\end{tabular}
\end{pmatrix}\,.
\end{align}
\end{subequations}

The couplings used in our calculations largely overlap with the bare couplings of the $N_\mathrm{f}=2+1$ flavour lattice QCD ensembles generated by the CLS consortium~\cite{Bruno:2014jqa,Bruno:2016plf,Bali:2016umi,Mohler:2017wnb}. These large-volume gauge field configuration ensembles with open and periodic boundary conditions emerge from large-scale numerical simulations with the same lattice action and at almost the same range of lattice spacings as in the present work. Since they are designed for a variety of phenomenological lattice QCD applications, for the sake of completeness and for prospective use in future calculations involving the renormalised and $\rmO(a)$ improved vector currents, we also interpolate (respectively, in the case of $\beta=3.85$, slightly extrapolate) our results to the corresponding values of $\beta=6/g_0^2$ employed there. The corresponding estimates for $Z_\mathrm{V}$ (resorting to the parameterisation~(\ref{eq:interpolation_ZV}) from Section~\ref{sec:renormalisation}), $c_\mathrm{V}$ and $c_{\tilde{\mathrm{V}}}$ are given in Table~\ref{tab:cls_range}.
\begin{table}[h]
	\centering
	\renewcommand{\arraystretch}{1.25}
	\setlength{\tabcolsep}{3pt}
	\begin{tabular}{llll}
		\toprule
		$\beta$ & $Z_\mathrm{V}$   & $c_\mathrm{V}$   & $c_{\tilde{\mathrm{V}}}$   \\
		\midrule
		3.34 & 0.69862(17)      & $-$0.346(17)       & 0.201(9)                   \\
		3.4  & 0.71066(11)      & $-$0.299(12)       & 0.232(7)                   \\
		3.46 & 0.72147(9)       & $-$0.259(8)        & 0.259(5)                   \\
		3.55 & 0.73571(10)      & $-$0.209(6)        & 0.294(4)                   \\
		3.7  & 0.75517(7)       & $-$0.147(8)        & 0.340(6)                   \\
		3.85 & 0.77058(11)      & $-$0.105(15)       & 0.374(12)                  \\
		\bottomrule
	\end{tabular}
	\caption{$Z_\mathrm{V}$, $c_\mathrm{V}$ and $c_{\tilde{\mathrm{V}}}$ results for the inverse gauge couplings $\beta=6/g_0^2$ of the $N_\mathrm{f}=2+1$ CLS simulations~\cite{Bruno:2014jqa,Bruno:2016plf,Bali:2016umi,Mohler:2017wnb}. The errors are the statistical uncertainties propagated from the interpolation formulae (\ref{eq:interpolation_ZV}) and (\ref{eq:interpolation_cV}), except for $\beta=3.85$, which lies slightly outside of the coupling range covered in the present investigation, where we add $50$\% of the size of the statistical error in quadrature as a systematic uncertainty accounting for this.}
	\label{tab:cls_range}
\end{table}
\subsection{Consistency checks and scaling tests}
We close this section with a discussion of a few consistency checks and scaling tests to demonstrate the vanishing of possible $\rmO(a)$ ambiguities.

First of all, as for the same check of volume dependence as for $Z_\mathrm{V}$ in Subsection~\ref{subsec:lcp_violation}, we observe for the improvement coefficients a qualitatively very similar picture that (on the level of larger statistical errors) even points to a statistically insignificant $L/a$-dependence. Hence, the general conclusion drawn there also applies here.
Furthermore, as an additional validation of our parameterising fit ansaetze in eqs.~(\ref{eq:interpolation_cV}), we evaluate the improvement conditions in the weak-coupling regime, i.e., on a lattice of size $16^3\times24$ with $g_0^2=0.75$, and at the chiral point. $Z_\mathrm{A}$ and $Z_\mathrm{V}^\mathrm{f}$ are extracted along the lines of \cite{Bulava:2016ktf} and Section~\ref{sec:renormalisation} of this this work, respectively, while $c_\mathrm{A}$ is set to the one-loop prediction of \cite{Aoki:1998qd}. As already mentioned before, the resulting numbers are also displayed in the left part of Figure~\ref{fig:cV_interpolations}. It is important to note, though, that this lattice has a distinctly smaller volume and does not lie on the LCP underlying the computations in the present work. Hence, although we do not expect these data points to perfectly comply with our interpolation formulae, we nevertheless see them as qualitatively very convincing support that our results in the regime of non-perturbative couplings, obtained via the Ward identity method, approach the perturbative predictions towards smaller bare couplings and that an ansatz monotonically connecting these two regimes is perfectly justified.

Next, in the top right part of Figure~\ref{fig:cV_interpolations}, we depict the outcome of a scaling analysis of different determinations of the conserved vector current improvement coefficient, $c_{\tilde{\mathrm{V}}}$.
In the following discussion we exclude the data points at the coarsest lattice spacing, corresponding to $\beta=3.3$ and outside of the CLS region, in order to describe the data with a reasonable functional form.
The deviation of the difference of the two determinations $c_{\tilde{\mathrm{V}}}^{\lbrace T/3\rbrace}$ and $c_{\tilde{\mathrm{V}}, \mathrm{alt}}^{\lbrace T/3,Z_\mathrm{V}^\chi\rbrace}$ from zero can be described by a sum of terms quadratic and cubic in the lattice spacing. This is expected, as both are based on the same correlation functions and only differ by how the first term in the improvement condition is renormalised. The difference of the two determinations $c_{\tilde{\mathrm{V}}, \mathrm{alt}}^{\lbrace T/4,Z_\mathrm{V}^\chi\rbrace}$ and $c_{\tilde{\mathrm{V}}, \mathrm{alt}}^{\lbrace T/3,Z_\mathrm{V}^\chi\rbrace}$ exhibits much smaller corrections to zero that can reliably be fitted by a purely quadratic term in the lattice spacing. Since the operator placement $t_1$ of $A_0$ in the bulk is the only difference between the two, this is in line with our expectation that no dominant higher-order ambiguity is present here. The scaling of the difference of the result by G\'{e}rardin et al.\ \cite{Gerardin:2018kpy} and our preferred determination is also well described by a quadratic ansatz in the lattice spacing of the form $0+c\cdot a^2$. In essence, intrinsic ambiguities among different definitions of $c_{\tilde{\mathrm{V}}}$-estimators, and even among completely independent determinations, vanish with a rate of at least $\mathrm{O}(a)$ towards the continuum limit as theoretically expected.

Similarly, the scaling analysis of different calculations of the local vector current improvement coefficient, $c_{\mathrm{V}}$, is illustrated in the bottom right part of Figure~\ref{fig:cV_interpolations}. Also here we exclude the data points at the coarsest lattice spacing. The deviation of the difference of our two determinations $c_{\mathrm{V}}^{\lbrace T/3\rbrace}$ and $c_{\mathrm{V}, \mathrm{alt}}^{\lbrace T/3,Z_\mathrm{V}^\chi\rbrace}$ from zero can again be fitted well by a sum of terms quadratic and cubic in the lattice spacing. The difference of $c_{\mathrm{V}, \mathrm{alt}}^{\lbrace T/4,Z_\mathrm{V}^\chi\rbrace}$ and $c_{\mathrm{V}, \mathrm{alt}}^{\lbrace T/3,Z_\mathrm{V}^\chi\rbrace}$, which only differ by the bulk position of the $A_0$ operators in the $k_{\mathrm{A}\mathrm{A}}^\mathrm{I}$- and $\tilde{k}_{\mathrm{P}\mathrm{A}}$-correlators, can be described by a term quadratic in the lattice spacing. For the difference of the result from \cite{Gerardin:2018kpy} and our preferred determination, we find leading corrections that are linear in the lattice spacing, also in line with the theoretical expectation.

\section{Conclusions}
\label{sec:concl}
In the present investigation we have determined the $\rmO(a)$ improvement coefficients of the local and conserved (i.e., point-split) versions of the lattice QCD vector current, as well as the $Z$-factor fixing its normalisation, for the theory discretised \`a la Wilson. Our non-perturbative calculations are performed in the framework of lattice QCD with ${\Nf=3}$ flavours of mass-degenerate, non-perturbatively $\rmO(a)$ improved~\cite{Bulava:2013cta} Wilson-clover sea quarks and tree-level Symanzik-improved~\cite{Luscher:1984xn} gluons. The underlying methodology is derived from massive chiral Ward identities, which are formulated and evaluated through numerical simulations in small, fixed physical volume of spatial extent $L \approx 1.2\,\fm$ and $T/L \approx 3/2$, with Schrödinger functional boundary conditions and an unitary setup of valence quark masses equal to those of the sea quarks. In this way we are able to approach the chiral and continuum limits, while staying on a line of constant physics (LCP) when the Ward identities are imposed to be restored up to $\rmO(a^2)$ at finite lattice spacing. Since the Ward identities are valid as operator relations in all sectors with fixed topology, the correlation functions involved have been projected to the sector of zero topological charge in order to ensure a reliable statistical error estimation from Monte Carlo simulation data facing topology freezing. 

Our main findings are the parameters of the interpolation formulae (\ref{eq:interpolation_ZV}) and (\ref{eq:interpolation_cV}) for $Z_{\mathrm{V}}$ and $c_{\mathrm{V}}$, $c_{\tilde{\mathrm{V}}}$, respectively, which hold for the range of bare couplings $1.55\lesssim g_0^2\lesssim 1.85$, i.e., lattice spacings $0.042\,\fm \lesssim a\lesssim 0.105\,\fm$. It also covers the $g_0^2$-values of the large-volume $N_\mathrm{f}=2+1$ simulations by CLS~\cite{Bruno:2014jqa,Bruno:2016plf,Bali:2016umi,Mohler:2017wnb}, and the corresponding results at these couplings are collected in Table~\ref{tab:cls_range}.\footnote{Only in case of $\beta=6/g_0^2=3.85$, which lies slightly outside the range of our computations, a very mild and thus safe extrapolation was necessary, see Subsection~\ref{subsec:improvement_res}.} Note that in this strong-coupling region and for the improvement coefficients we observe sizable non-perturbative corrections to the one-loop (resp.\ tree-level) prediction, by contrast to the outcome of the study~\cite{Gerardin:2018kpy} employing a different improvement condition. Since the uncertainties on $c_{\mathrm{V}}$ and $c_{\tilde{\mathrm{V}}}$ receive a significant contribution from the (independent) statistical error on the ratio $Z_{\mathrm{A}}^2/Z_{\mathrm{V}}^{\phantom{2}}$ and may suffer from systematics due to unbalanced higher-order ambiguities between numerator and denominator, we have advocated a prescription where this ratio is replaced by $Z_{\mathrm{A}}^2/Z_{\mathrm{V}}^2$, very accurately know from Ref.~\cite{DallaBrida:2018tpn}, such that these effects are largely reduced. At this point, however, we would like to highlight again that, in the very spirit of the generic LCP idea, the exact choice of $c$-coefficients and $Z$-factors from our results is not decisive from the theoretical perspective as long as one stays within the set of numbers for a given definition and does not switch between different definitions when changing the bare coupling. Otherwise cutoff effects are no longer guaranteed to vanish smoothly at a rate $\propto a^2$.

Thanks to the constant physics condition, our results for the improvement coefficients and the normalisation factor exhibit a smooth dependence on the (squared) bare gauge coupling, monotonically linking the non-perturbative domain, where they typically are relevant for further applications in various contexts of hadronic physics, with the perturbative regime. Moreover, to demonstrate the negligibility of possibly uncontrolled systematic effects, we have demonstrated explicitly that variations in the actual numerical implementation of the Ward identity based improvement and normalisation conditions only give rise to ambiguities of expected higher order (viz., at least $\rmO(a)$ for $c_{\mathrm{V}}$ and $c_{\tilde{\mathrm{V}}}$, and $\rmO(a^2)$ for $Z_{\mathrm{V}}$) that uniformly disappear as $g_0^2\to 0$.

We also find overall consistency of our results with those obtained in Refs.~\cite{DallaBrida:2018tpn,Gerardin:2018kpy} by similar methods, though grounded on different ensembles and (in case of~\cite{Gerardin:2018kpy}) not strictly obeying the condition of constant physics. While for $Z_{\mathrm{V}}$ the respective $g_0^2$-dependences fall quite close to each other, $c_{\mathrm{V}}$ and $c_{\tilde{\mathrm{V}}}$ clearly disagree quite substantially from~\cite{Gerardin:2018kpy} at couplings over the common range of finite lattice spacings $0.04\,\Fm\lesssim a\lesssim0.1\,\Fm$, even though these differences are perfectly in line with $\mathrm{O}\big(a^{(n\ge 1)}\big)$ ambiguities that as a consequence of completely independent improvement and renormalisation conditions are inherently unavoidable but vanish towards weaker couplings. In particular, together with the result for $c_\mathrm{V}$ in Ref.~\cite{Gerardin:2018kpy}, the outcome of our determination indicates that its one-loop perturbative prediction is compatible with both results, but up to significantly different higher-order cutoff effects.  Therefore, it remains to be seen which set of estimates will in practice lead to a more pronounced impact of improvement in the scaling behaviour of quantities involving the vector current; this could, for instance, be quantitatively explored in a dedicated precision intermediate-volume scaling test similar to~\cite{Heitger:1999dw} or in future phenomenological studies.\footnote{Recall that since the improvement of spectral quantities only depends on the improvement of the action, the effect of the vector current improvement is expected to be most significant for its matrix elements (as they, e.g., enter decay constants and form factors).} As for the latter, interesting applications of the renormalised $\rmO(a)$ improved vector current and its matrix elements include computations of vector meson decay constants (such as, in the charm sector, the $D^\ast$ or $J/\psi$ decay constants), semi-leptonic decay form factors, the leading-order hadronic vacuum polarisation and light-by-light contributions to the muon's anomalous magnetic moment, as well as thermal correlators related to the di-lepton production rate in the quark-gluon plasma.

\begin{acknowledgement}%
We wish to thank Patrick Fritzsch, Antoine G\'{e}rardin, Simon Kuberski, Rainer Sommer and Christian Wittemeier for useful discussions. We are specifically grateful to Christian Wittemeier for his help in implementing further Schrödinger functional correlation functions into our measurement code at an early stage of this project, Carl Christian Köster and Patrick Fritzsch for their valuable contributions in extending the set of available ensembles, and to Rainer Sommer for valuable comments on an earlier version of this manuscript.
This work is supported by the Deutsche Forschungsgemeinschaft (DFG) through the Research Training Group {\em GRK 2149: Strong and Weak Interactions – from Hadrons to Dark Matter}. We gratefully acknowledge the computer resources provided by the IT Center of Münster University (PALMA II HPC cluster) and thank the WWU IT staff for support.

\end{acknowledgement}

\clearpage
\appendix

\section{Result tables} \label{app:tables}

\begin{table}[h]
	\centering
	\renewcommand{\arraystretch}{1.25}
	\setlength{\tabcolsep}{3pt}
	\begin{tabular}{cl >{\em}ll}
		\toprule
		ID   & $am$   & $Z_\mathrm{V}^\mathrm{f}$   & $Z_\mathrm{V}^\mathrm{k}$   \\
		\midrule
		A1k1 & $-$0.00287(61)           & 0.68925(34)                 & 0.69434(87)                 \\
		A1k3 & \phantom{$-$}0.00105(95) & 0.69027(51)                 & 0.69533(172)                \\
		A1k4 & $-$0.00119(33)           & 0.68914(24)                 & 0.69515(64)                 \\
		& \phantom{$-$}0.0         & 0.68964(28)                 & 0.69540(84)                 \\
		\midrule
		E1k1 & \phantom{$-$}0.00270(20) & 0.71524(15)                 & 0.71913(42)                 \\
		E1k2 & $-$0.00013(17)           & 0.71329(13)                 & 0.71704(48)                 \\
		& \phantom{$-$}0.0         & 0.71338(15)                 & 0.71714(45)                 \\
		\midrule
		B1k1 & \phantom{$-$}0.00552(20) & 0.73403(13)                 & 0.73668(32)                 \\
		B1k2 & \phantom{$-$}0.00435(28) & 0.73327(23)                 & 0.73501(73)                 \\
		B1k3 & \phantom{$-$}0.00157(18) & 0.73072(16)                 & 0.73511(54)                 \\
		B1k4 & $-$0.00056(16)           & 0.72980(14)                 & 0.73289(34)                 \\
		& \phantom{$-$}0.0         & 0.72999(14)                 & 0.73340(30)                 \\
		\midrule
		C1k1 & \phantom{$-$}0.01322(17) & 0.76345(13)                 & 0.76562(39)                 \\
		C1k2 & \phantom{$-$}0.00601(11) & 0.75713(12)                 & 0.75922(36)                 \\
		C1k3 & $-$0.00110(11)           & 0.75142(11)                 & 0.75309(21)                 \\
		& \phantom{$-$}0.0         & 0.75226(12)                 & 0.75404(21)                 \\
		\midrule
		D1k2 & \phantom{$-$}0.00073(15) & 0.76733(28)                 & 0.76923(62)                 \\
		D1k4 & $-$0.00007(3)            & 0.76676(5)                  & 0.76781(9)                  \\
		& \phantom{$-$}0.0         & 0.76681(5)                  & 0.76793(10)                 \\
		\bottomrule
	\end{tabular}
	\caption{Summary of results for the PCAC quark mass $am$ and the two determinations of the vector renormalization constant $Z_\mathrm{V}^\mathrm{f}$ and $Z_\mathrm{V}^\mathrm{k}$ on the individual ensembles and in the chiral limit. The errors for the individual ensembles are statistical, the ones in the chiral limit follow from the orthogonal distance regression procedure of Ref.~\cite{Boggs1989}. Our preferred determination $Z_\mathrm{V}^\mathrm{f}$ is emphasized in italic.}
	\label{tab:ZV_results}
\end{table}

\begin{table}[h]
	\centering
	\renewcommand{\arraystretch}{1.25}
	\setlength{\tabcolsep}{3pt}
	\begin{tabular}{lllllll >{\em}l}
		\toprule
		ID   & $am$      & $c_\mathrm{V}^{\lbrace T/4, Z_\mathrm{V}^\mathrm{f} \rbrace}$   & $c_\mathrm{V}^{\lbrace T/3, Z_\mathrm{V}^\mathrm{f} \rbrace}$   & $c_\mathrm{V}^{\lbrace T/4, Z_\mathrm{V}^\chi \rbrace}$   & $c_\mathrm{V}^{\lbrace T/3, Z_\mathrm{V}^\chi \rbrace}$   & $c_{\mathrm{V}, \mathrm{alt}}^{\lbrace T/4, Z_\mathrm{V}^\chi \rbrace}$   & $c_{\mathrm{V}, \mathrm{alt}}^{\lbrace T/3, Z_\mathrm{V}^\chi \rbrace}$   \\
		\midrule
		A1k1 & $-$0.00287(61)           & $-$0.259(32)                                                      & $-$0.235(26)                                                      & $-$0.487(58)                                                & $-$0.405(54)                                                & $-$0.478(35)                                                          & $-$0.456(28)                                                          \\
		A1k3 & \phantom{$-$}0.00105(95) & $-$0.240(37)                                                      & $-$0.184(30)                                                      & $-$0.504(61)                                                & $-$0.421(53)                                                & $-$0.426(35)                                                          & $-$0.373(30)                                                          \\
		A1k4 & $-$0.00119(33)           & $-$0.221(20)                                                      & $-$0.207(17)                                                      & $-$0.489(30)                                                & $-$0.389(27)                                                & $-$0.431(23)                                                          & $-$0.419(21)                                                          \\
		& \phantom{$-$}0.0         & $-$0.227(24)                                                      & $-$0.194(21)                                                      & $-$0.496(36)                                                & $-$0.403(33)                                                & $-$0.427(26)                                                          & $-$0.394(25)                                                          \\
		\midrule
		E1k1 & \phantom{$-$}0.00270(20) & $-$0.243(15)                                                      & $-$0.194(12)                                                      & $-$0.352(25)                                                & $-$0.284(24)                                                & $-$0.335(16)                                                          & $-$0.288(13)                                                          \\
		E1k2 & $-$0.00013(17)           & $-$0.238(16)                                                      & $-$0.171(12)                                                      & $-$0.333(28)                                                & $-$0.311(26)                                                & $-$0.356(16)                                                          & $-$0.291(13)                                                          \\
		& \phantom{$-$}0.0         & $-$0.238(15)                                                      & $-$0.172(11)                                                      & $-$0.334(27)                                                & $-$0.310(25)                                                & $-$0.355(16)                                                          & $-$0.291(13)                                                          \\
		\midrule
		B1k1 & \phantom{$-$}0.00552(20) & $-$0.186(14)                                                      & $-$0.169(12)                                                      & $-$0.229(20)                                                & $-$0.200(19)                                                & $-$0.194(15)                                                          & $-$0.177(12)                                                          \\
		B1k2 & \phantom{$-$}0.00435(28) & $-$0.169(25)                                                      & $-$0.156(19)                                                      & $-$0.206(37)                                                & $-$0.188(32)                                                & $-$0.190(26)                                                          & $-$0.177(18)                                                          \\
		B1k3 & \phantom{$-$}0.00157(18) & $-$0.221(18)                                                      & $-$0.188(13)                                                      & $-$0.264(28)                                                & $-$0.227(25)                                                & $-$0.273(17)                                                          & $-$0.240(14)                                                          \\
		B1k4 & $-$0.00056(16)           & $-$0.202(15)                                                      & $-$0.150(11)                                                      & $-$0.230(30)                                                & $-$0.226(25)                                                & $-$0.272(15)                                                          & $-$0.220(12)                                                          \\
		& \phantom{$-$}0.0         & $-$0.207(12)                                                      & $-$0.161(9)                                                       & $-$0.244(23)                                                & $-$0.227(19)                                                & $-$0.273(12)                                                          & $-$0.227(10)                                                          \\
		\midrule
		C1k1 & \phantom{$-$}0.01322(17) & $-$0.252(15)                                                      & $-$0.217(13)                                                      & $-$0.180(18)                                                & $-$0.166(16)                                                & $-$0.110(16)                                                          & $-$0.074(15)                                                          \\
		C1k2 & \phantom{$-$}0.00601(11) & $-$0.197(14)                                                      & $-$0.180(11)                                                      & $-$0.173(20)                                                & $-$0.143(19)                                                & $-$0.131(16)                                                          & $-$0.114(13)                                                          \\
		C1k3 & $-$0.00110(11)           & $-$0.144(18)                                                      & $-$0.140(14)                                                      & $-$0.193(39)                                                & $-$0.147(38)                                                & $-$0.167(19)                                                          & $-$0.163(16)                                                          \\
		& \phantom{$-$}0.0         & $-$0.152(14)                                                      & $-$0.146(11)                                                      & $-$0.181(26)                                                & $-$0.137(25)                                                & $-$0.159(16)                                                          & $-$0.154(14)                                                          \\
		\midrule
		D1k2 & \phantom{$-$}0.00073(15) & $-$0.118(24)                                                      & $-$0.127(15)                                                      & $-$0.135(49)                                                & $-$0.113(40)                                                & $-$0.107(25)                                                          & $-$0.117(17)                                                          \\
		D1k4 & $-$0.00007(3)            & $-$0.136(11)                                                      & $-$0.115(10)                                                      & $-$0.118(20)                                                & $-$0.114(18)                                                & $-$0.138(14)                                                          & $-$0.117(13)                                                          \\
		& \phantom{$-$}0.0         & $-$0.135(10)                                                      & $-$0.116(10)                                                      & $-$0.120(19)                                                & $-$0.114(17)                                                & $-$0.136(13)                                                          & $-$0.117(13)                                                          \\
		\bottomrule
	\end{tabular}
	\caption{Summary of results for $am$ and different determinations of the local vector current improvement coefficient $c_\mathrm{V}$ on the individual ensembles and in the chiral limit. The errors for the individual ensembles are statistical, the ones in the chiral limit follow from the orthogonal distance regression procedure of Ref.~\cite{Boggs1989}. Our preferred determination $c_{\mathrm{V}, \mathrm{alt}}^{\lbrace T/3, Z_\mathrm{V}^\chi \rbrace}$ is emphasized in italic.}
	\label{tab:cV_results}
\end{table}

\begin{table}[h]
	\centering
	\renewcommand{\arraystretch}{1.25}
	\setlength{\tabcolsep}{3pt}
	\begin{tabular}{lllll >{\em}l}
		\toprule
		ID   & $am$      & $c_{\tilde{\mathrm{V}}}^{\lbrace T/4 \rbrace}$   & $c_{\tilde{\mathrm{V}}}^{\lbrace T/3 \rbrace}$   & $c_{\tilde{\mathrm{V}}, \mathrm{alt}}^{\lbrace T/4, Z_\mathrm{V}^\chi \rbrace}$   & $c_{\tilde{\mathrm{V}}, \mathrm{alt}}^{\lbrace T/3, Z_\mathrm{V}^\chi \rbrace}$   \\
		\midrule
		A1k1 & $-$0.00287(61)           & 0.269(22)                                        & 0.285(18)                                        & 0.116(24)                                                                     & 0.132(20)                                                                     \\
		A1k3 & \phantom{$-$}0.00105(95) & 0.277(25)                                        & 0.315(21)                                        & 0.148(24)                                                                     & 0.185(20)                                                                     \\
		A1k4 & $-$0.00119(33)           & 0.290(14)                                        & 0.300(11)                                        & 0.145(16)                                                                     & 0.154(14)                                                                     \\
		& \phantom{$-$}0.0         & 0.285(16)                                        & 0.308(14)                                        & 0.147(17)                                                                     & 0.170(17)                                                                     \\
		\midrule
		E1k1 & \phantom{$-$}0.00270(20) & 0.271(11)                                        & 0.306(9)                                         & 0.204(12)                                                                     & 0.238(10)                                                                     \\
		E1k2 & $-$0.00013(17)           & 0.276(11)                                        & 0.324(9)                                         & 0.193(12)                                                                     & 0.240(10)                                                                     \\
		& \phantom{$-$}0.0         & 0.276(11)                                        & 0.324(8)                                         & 0.194(11)                                                                     & 0.240(9)                                                                      \\
		\midrule
		B1k1 & \phantom{$-$}0.00552(20) & 0.309(10)                                        & 0.321(9)                                         & 0.304(11)                                                                     & 0.317(9)                                                                      \\
		B1k2 & \phantom{$-$}0.00435(28) & 0.324(19)                                        & 0.334(14)                                        & 0.309(19)                                                                     & 0.318(14)                                                                     \\
		B1k3 & \phantom{$-$}0.00157(18) & 0.283(12)                                        & 0.307(10)                                        & 0.245(13)                                                                     & 0.268(10)                                                                     \\
		B1k4 & $-$0.00056(16)           & 0.302(11)                                        & 0.339(8)                                         & 0.252(11)                                                                     & 0.289(9)                                                                      \\
		& \phantom{$-$}0.0         & 0.296(9)                                         & 0.330(7)                                         & 0.248(9)                                                                      & 0.282(8)                                                                      \\
		\midrule
		C1k1 & \phantom{$-$}0.01322(17) & 0.256(11)                                        & 0.283(10)                                        & 0.364(12)                                                                     & 0.392(12)                                                                     \\
		C1k2 & \phantom{$-$}0.00601(11) & 0.299(11)                                        & 0.312(9)                                         & 0.349(12)                                                                     & 0.362(10)                                                                     \\
		C1k3 & $-$0.00110(11)           & 0.343(13)                                        & 0.346(11)                                        & 0.328(15)                                                                     & 0.330(12)                                                                     \\
		& \phantom{$-$}0.0         & 0.336(11)                                        & 0.340(9)                                         & 0.332(13)                                                                     & 0.336(10)                                                                     \\
		\midrule
		D1k2 & \phantom{$-$}0.00073(15) & 0.361(18)                                        & 0.354(12)                                        & 0.369(21)                                                                     & 0.362(16)                                                                     \\
		D1k4 & $-$0.00007(3)            & 0.348(8)                                         & 0.364(8)                                         & 0.346(10)                                                                     & 0.362(10)                                                                     \\
		& \phantom{$-$}0.0         & 0.349(8)                                         & 0.363(7)                                         & 0.348(10)                                                                     & 0.362(10)                                                                     \\
		\bottomrule
	\end{tabular}
	\caption{Summary of results for $am$ and different determinations of the conserved vector current improvement coefficient $c_{\tilde{\mathrm{V}}}$ on the individual ensembles and in the chiral limit. The errors for the individual ensembles are statistical, the ones in the chiral limit follow from the orthogonal distance regression procedure of Ref.~\cite{Boggs1989}. Our preferred determination $c_{\tilde{\mathrm{V}}, \mathrm{alt}}^{\lbrace T/3, Z_\mathrm{V}^\chi \rbrace}$ is emphasized in italic.}
	\label{tab:cV_tilde_results}
\end{table}
\clearpage
\section{Schrödinger functional correlation functions}
\label{sec:sf_correlation_functions}
The Appendix summarises the definitions of the Schrödinger functional correlation functions employed in this work.
\begin{align}
f_\mathrm{P}(x_0) &=-\frac{1}{2}\frac{a^9}{L^3}\sum_{\mathbf{x}}
\big\langle P^a(x_0,\mathbf{x})\mathcal{O}^a \big\rangle\,,\\
f_\mathrm{A}(x_0) &=-\frac{1}{2}\frac{a^9}{L^3}\sum_{\mathbf{x}}
\big\langle A_0^a(x_0,\mathbf{x})\mathcal{O}^a \big\rangle\,,\\
k_\mathrm{V}(x_0) &=-\frac{1}{6}\frac{a^9}{L^3}\sum_{\mathbf{x}}
\big\langle V_k^a(x_0,\mathbf{x})\mathcal{Q}_k^a \big\rangle\,,\\
k_{\tilde{\mathrm{V}}}(x_0) &=-\frac{1}{6}\frac{a^9}{L^3}\sum_{\mathbf{x}}
\big\langle \tilde{V}_k^a(x_0,\mathbf{x})\mathcal{Q}_k^a \big\rangle\,,\\
k_\mathrm{T}(x_0) &=-\frac{1}{6}\frac{a^9}{L^3}\sum_{\mathbf{x}}
\big\langle T_{k0}^a(x_0,\mathbf{x})\mathcal{Q}_k^a \big\rangle\,,\\
k_{\mathrm{A}\mathrm{A}}(y_0, x_0) &=-\frac{1}{6}\frac{a^9}{L^3}\epsilon^{abc}\sum_{\mathbf{x},\mathbf{y}}
\big\langle A_0^a(y_0,\mathbf{y})A_k^b(x_0,\mathbf{x})\mathcal{Q}_k^c \big\rangle\,,\\
k_{\mathrm{P}\mathrm{A}}(y_0, x_0) &=-\frac{1}{6}\frac{a^9}{L^3}\epsilon^{abc}\sum_{\mathbf{x},\mathbf{y}}
\big\langle P^a(y_0,\mathbf{y})A_k^b(x_0,\mathbf{x})\mathcal{Q}_k^c \big\rangle\,,\\
k_{\mathrm{A}\mathrm{A}}^\mathrm{I}(y_0, x_0)&=k_{\mathrm{A}\mathrm{A}}(y_0, x_0)+c_\mathrm{A}\tilde{\partial}_{y_0}k_{\mathrm{P}\mathrm{A}}(y_0, x_0)\,,\\
\tilde{k}_{\mathrm{P}\mathrm{A}}(t_1, t_2, x_0) &=-\frac{1}{6}\frac{a^9}{L^3}\epsilon^{abc}\sum_{y_0=t_1}^{t_2}w(y_0)\sum_{\mathbf{x},\mathbf{y}}
\big\langle P^a(y_0,\mathbf{y})A_k^b(x_0,\mathbf{x})\mathcal{Q}_k^c \big\rangle\,,\\
w(y_0)&=\begin{cases}
\tfrac{1}{2}\,, & y_0\in\lbrace t_1, t_2 \rbrace\\
1\,, & \text{otherwise}
\end{cases} \,.
\nonumber
\end{align}

\clearpage
\small
\addcontentsline{toc}{section}{References}
\bibliographystyle{JHEP}
\bibliography{bib}

\end{document}